\documentclass[sigconf]{acmart}

\AtBeginDocument{%
  }

\usepackage{booktabs}
\usepackage{graphicx}
\usepackage{balance}
\usepackage{xcolor}
\usepackage{soul}

\setcopyright{none}
\acmConference[Preprint]{Preprint}{2026}{}
\acmYear{2026}
\copyrightyear{2026}
\acmDOI{}
\acmISBN{}

\begin{document}

% \title{Same Stories, Different Journeys: From Social Comparison to Sensemaking in AI-Mediated Peer Career Exploration}
\title{Same Stories, Different Journeys: Exploring Persona-Grounded Conversational Agents for Supporting Career Exploration with Peers' Posts}

\author{Pengping Tan}
\affiliation{%
  \institution{Sun Yat-sen University}
  \city{Guangzhou}
  \country{China}
}
\email{tanpp5@mail2.sysu.edu.cn}

\author{Baoquan Zhao}
\affiliation{%
  \institution{Sun Yat-sen University}
  \city{Zhuhai}
  \country{China}
}
\email{zhaobaoquan@mail.sysu.edu.cn}

\author{Shuai Ma}
\affiliation{%
      \institution{Institute of Software, Chinese Academy of Sciences}
  \city{Beijing}
  \country{China}
}
\email{mashuai@iscas.ac.cn}

\author{Zhenhui Peng\textsuperscript{*}}
\affiliation{%
  \institution{Sun Yat-sen University}
  \city{Zhuhai}
  \country{China}
}
\email{pengzhh29@mail.sysu.edu.cn}

\renewcommand{\shortauthors}{Tan et al.}

%% ============================================================
%% ABSTRACT
%% ============================================================
\begin{abstract}

Young job seekers frequently explore their career possibilities by browsing peers' posts that share job-seeking experiences. However, static browsing requires them to reconstruct fragmented cases and privately judge what others' experiences mean for themselves, sometimes intensifying anxiety through upward social comparison. In this paper, we examine how transforming these posts into persona-grounded conversations reshapes this sensemaking process. We developed JobMate, a prototype featuring agents that have personas built upon peers' posts and follow the self-determination theory to converse with users. In a between-subjects comparative study with 24 participants, RedNote browsing exposed diverse trajectories but left reconstruction and comparison largely to users, whereas JobMate supported case selection and continued questioning. The conversations further prompted users to articulate previously implicit constraints and accept, challenge, or revise the agent's interpretations. We discuss design implications for combining authentic peer experiences with generative AI in career exploration.

%% 年轻求职者经常通过浏览同伴分享求职经历的帖子来探索自己的职业可能性。然而，静态浏览要求他们重构碎片化的案例，并在内心判断他人的经历对自己意味着什么；向上的社会比较有时还会加剧焦虑。本文考察将这些帖子转化为基于 persona 的对话，如何改变这一意义建构过程。我们开发了 JobMate，一个将小红书上的真实帖子组织为可比较的 personas，并支持来源可见、由自我决定理论引导且结合检索增强生成对话的研究 probe。在一项包含 24 名参与者的组间比较研究中，小红书浏览让用户接触到多样的职业轨迹，但主要由用户自行重构和比较案例；JobMate 则支持案例选择和持续追问。对话还促使用户说出此前保持内隐的个人限制，并接受、质疑或修订 agent 的解释。我们据此讨论在职业探索中结合真实同伴经历与生成式 AI 的设计启示。
  \end{abstract}




\begin{teaserfigure}
  \centering
  \includegraphics[width=\textwidth]{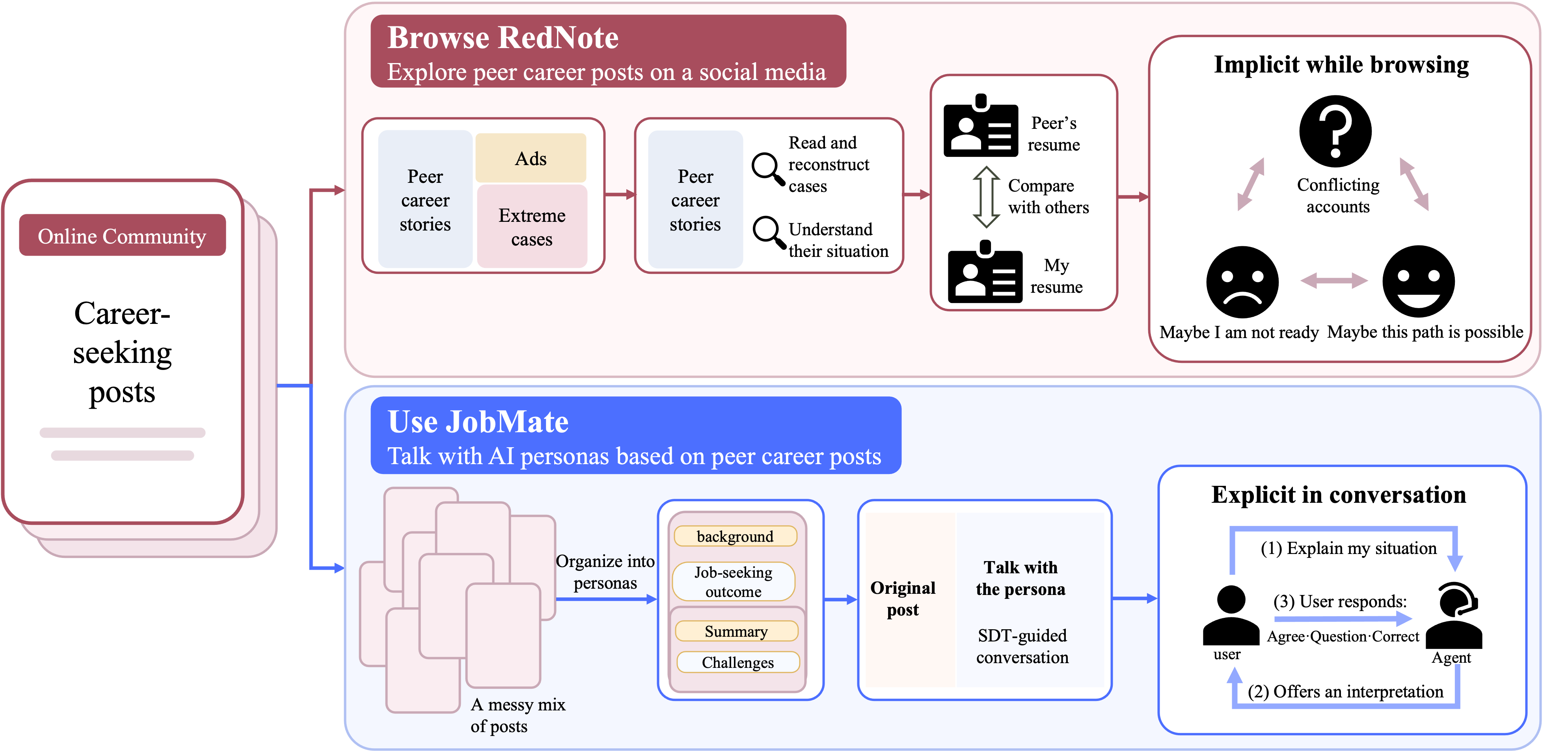}
  \caption{Overview of the two ways of exploring peer career posts in our study. When browsing RedNote (top), users find useful stories, work out each peer's situation, and compare it with their own; much of this interpretation stays implicit. In JobMate (bottom), posts are organized into personas. Users choose a persona, view the original post, explain their own situation in conversation, and agree with, question, or correct the agent's interpretation.}
  \Description{A three-stage diagram from a crowded collection of social-media posts to structured persona cards to a chat window, with simple illustrations of a stressed versus relaxed student.}
  \label{fig:overview}
\end{teaserfigure}

\maketitle

%% ============================================================
%% 1. INTRODUCTION —— 从"系统介绍"变为"科学问题驱动"
%% ============================================================
\section{Introduction}

Social media has become an important venue for young people to explore careers~\cite{Mowbray2021,Jiang2019CareerExploration}. Unlike official job descriptions or general career advice, social platforms expose job seekers to concrete accounts of how ordinary people with recognizable backgrounds navigated career choices, encountered difficulties, and arrived at particular outcomes. For young people with limited first-hand work experience, these peer-authored trajectories provide more than occupational information: they offer concrete cases through which users can imagine possible future selves and assess whether unfamiliar career paths might also be attainable for them~\cite{Corvite2024,Markus1986,Lent1994SCCT}. 
%% 社交媒体已经成为年轻人探索职业的重要场所~\cite{Mowbray2021}。不同于正式的职位描述或一般性的职业建议，社交平台让求职者接触到具体的经历叙述，了解具有可识别背景的普通人如何应对职业选择、遭遇困难，并最终走向特定的结果。对于缺少一手工作经验的年轻人而言，这些由同伴讲述的职业轨迹所提供的不只是职业信息；它们还构成了具体案例，使用户能够想象可能的未来自我，并判断陌生的职业路径对自己而言是否也可以实现~\cite{Corvite2024,Markus1986}。

Yet using these stories for career exploration requires more than simply understanding what happened to someone else. Job seekers must decide what another person's experience means for their own situation. This process involves two forms of interpretive work. First, career experiences are distributed across heterogeneous posts and comments, requiring users to locate relevant cases and reconstruct each author's background, challenges, and trajectory before judging whether the case is useful~\cite{Wu2025ComViewer,Li2022PlanHelper}. Second, even after understanding a peer's experience, users must determine whether and how it applies to themselves. Such interpretation often involves social comparison: a peer's qualifications, choices, and outcomes can become reference points against which users evaluate their own prospects. Prior work has associated career-related social comparison with frustration, job-search pressure, and employment anxiety~\cite{Fukubayashi2021,Zhang2024Pressure,Luo2024}. Thus, the same specificity that makes peer stories valuable also makes them difficult to use. Users are not merely extracting information from others' experiences; they are continually translating those experiences into judgments about themselves.
%% 然而，利用这些故事进行职业探索，并不只是理解别人经历了什么。求职者还必须判断，他人的经历对自己的处境意味着什么。这个过程包含两类解释工作。第一，职业经历分散在形式各异的帖子和评论中，用户需要找到相关案例，并重构每位作者的背景、困难和职业轨迹，之后才能判断该案例是否有用~\cite{Wu2025ComViewer,Li2022PlanHelper}。第二，即使理解了某位同伴的经历，用户仍需判断这段经历是否适用于自己，以及应当如何应用。这种解释不可避免地会引发社会比较：同伴的资历、选择和结果可能成为参照点，用户据此评估自己的职业前景。已有研究将职业相关的社会比较与挫败感、求职压力和就业焦虑联系起来~\cite{Fukubayashi2021,Zhang2024Pressure,Luo2024}。因此，使同伴故事具有价值的具体性，也让它们难以被使用。用户不只是在从他人的经历中提取信息，也在不断把这些经历转化为对自己的判断。

Importantly, social-media browsing leaves much of this translation private and implicit. Users can read another person's story, compare themselves with that person, and form a tentative judgment about what they lack or what might suit them, but the post itself cannot respond to that judgment. When a comparison feels inappropriate, when important context is missing, or when a user is unsure why another person's experience should apply to them, there is often no specific interlocutor with whom that interpretation can be examined. Conversational interaction creates a different possibility. Rather than only presenting another person's experience, an agent can ask users to articulate their own background, constraints, and concerns, offer an interpretation in return, and allow users to accept, challenge, or revise that interpretation. In this way, dialogue can potentially transform peer comparison from an internal judgment into an explicit and negotiable process of sensemaking. This distinction motivates us to examine not simply whether conversational agents provide better career information, but how they reshape the process through which users relate others' experiences to themselves. The later system design explicitly uses conversation to surface constraints and make interpretations available for acceptance, contestation, and revision.
%% 重要的是，社交媒体浏览使这一转化过程在很大程度上发生于个体内部，并保持内隐。用户可以阅读他人的故事、将自己与对方比较，并初步判断自己缺少什么或可能适合什么，但帖子本身无法回应这种判断。当用户认为某种比较并不恰当、重要背景有所缺失，或者不确定他人的经历为什么适用于自己时，通常没有一个明确的对话对象可以与其共同检视这种解释。对话式交互创造了另一种可能。Agent 不只是呈现他人的经历，还可以要求用户说出自己的背景、限制和顾虑，继而给出解释，并允许用户接受、质疑或修订这种解释。通过这种方式，对话有可能把同伴比较从一种内在判断转化为明确且可协商的意义建构过程。基于这一区别，我们考察的并不只是对话式 agent 能否提供更好的职业信息，而是它如何改变用户把他人经历关联到自身的过程。后文的系统设计明确利用对话来呈现用户的限制，并使系统给出的解释可以被接受、反驳和修订。

Existing conversational systems offer important foundations but do not fully capture this interaction. Career agents are commonly designed as normative counselors that guide decision-making~\cite{Gati2021CareerDecisions,DSilva2020CareerChatbot,Lee2021CASExplorer,Nguyen2022ITCareerBot,Han2025SDT,Du2024CareerSim}, or as projections of the user's own future self that support reflection and future-self continuity~\cite{Jeon2025LettersFutureSelf,Pataranutaporn2024FutureYou}. Persona-based agents introduce another perspective into conversation, but these personas are often fictional experts, generated stances, or simulated characters~\cite{Liu2025PersonaFlow,Park2023GenerativeAgents}. Even systems grounded in community-generated content have primarily focused on helping users retrieve, organize, or understand information~\cite{Wu2026ConSearcher}. What remains less understood is what happens when the conversational counterpart is grounded in the lived experience of a specific real peer: the agent does not merely provide information about a career path, but interprets the user from the standpoint of an identifiable peer experience, while the user can inspect, question, and respond to that interpretation. This changes the direction of sensemaking. Users interpret a peer's experience. In turn, an agent grounded in that experience offers an interpretation of the user's own situation, and the user can respond to it.
%% 现有的对话系统提供了重要基础，却没有完全涵盖这种交互。职业 agent 通常被设计为引导用户进行决策的规范性咨询者~\cite{Han2025SDT,Du2024CareerSim}，或被设计为用户未来自我的投射，以支持反思和未来自我连续性~\cite{Jeon2025LettersFutureSelf,Pataranutaporn2024FutureYou}。基于 persona 的 agent 为对话引入了另一个视角，但这些 persona 往往是虚构的专家、生成的立场或模拟的人物~\cite{Liu2025PersonaFlow,Park2023GenerativeAgents}。即使是基于社区生成内容的系统，也主要关注帮助用户检索、组织或理解信息~\cite{Wu2026ConSearcher}。我们仍不充分了解的是：当对话对象基于某位真实同伴的亲身经历时，会发生什么。Agent 不只是提供有关某条职业路径的信息，还会站在一段可识别的同伴经历的立场上解释用户；与此同时，用户可以检查、质疑并回应这种解释。这种交互引入了一种独特的意义建构方向：用户不只是在理解同伴的经历，也会接收并回应 agent 基于该经历对用户自身处境作出的解释。

We therefore ask: \emph{How does transforming peer career experiences from browsable social-media posts into persona-grounded conversational encounters reshape how job seekers make sense of those experiences in relation to themselves?} To investigate this question, we first conducted a formative study combining an online survey ($N{=}64$) and semi-structured interviews ($N{=}8$) to understand how young job seekers use social media and AI for career exploration. Participants valued peer experiences that were credible, comparable, and relevant to their own backgrounds, but described substantial cognitive and emotional work in filtering fragmented content, reconstructing suitable cases, and managing social comparison. Based on these findings, we developed JobMate, a research probe that transforms authentic RedNote career posts into structured personas. JobMate presents comparable information about each peer's background, outcome, experience, and challenges, supports dialogue guided by Self-Determination Theory (SDT), and uses retrieval-augmented generation (RAG) to incorporate related peer experiences and practical information while keeping the original post available for inspection.
%% 因此，我们提出以下研究问题：\emph{将同伴职业经历从可浏览的社交媒体帖子转化为基于 persona 的对话式相遇，会如何改变求职者联系自身来理解这些经历的方式？} 为研究这一问题，我们首先开展了一项结合在线问卷（$N{=}64$）与半结构化访谈（$N{=}8$）的形成性研究，以理解年轻求职者如何使用社交媒体和 AI 进行职业探索。参与者重视可信、可比较且与自身背景相关的同伴经历，但也表示，筛选碎片化内容、重构合适的案例以及应对社会比较，需要付出大量认知和情绪劳动。这些发现促使我们构建 JobMate：一个将小红书上的真实职业帖子转化为结构化的 persona 的研究 probe。JobMate 呈现关于每位同伴的背景、结果、经历和困难的可比较信息，支持由自我决定理论（Self-Determination Theory，SDT）引导的对话，并使用检索增强生成（retrieval-augmented generation，RAG）将相关同伴经历和实用信息融入对话，同时保留原帖以供用户核查。

We then conducted a between-subjects comparative study with 24 university students, comparing JobMate with native RedNote browsing. We use this comparison to examine two ways of engaging with peer-authored career experiences. We do not use it to isolate the effect of any single interface component. We observed that native browsing exposed participants to a broad range of trajectories but required them to reconstruct cases and make comparisons largely on their own. JobMate reorganized exploration around selectable peer cases and continued questioning, while conversation prompted users to articulate personal constraints and criteria that had previously remained implicit. In particular, participants could respond to the agent's interpretations of their abilities, experiences, and career fit by accepting, challenging, or revising them. When persona matching was unclear, desired career paths were absent from the source corpus, or responses remained general, participants had less reason to select a persona or continue the conversation.
%% 随后，我们开展了一项包含 24 名大学生的组间比较研究，将 JobMate 与原生小红书浏览进行比较。我们用这一比较来考察用户参与同伴职业经历的两种方式，而不是分离任何单一界面组件的效果。我们观察到，原生浏览使参与者接触到广泛的职业轨迹，但他们在很大程度上需要自行重构案例并进行比较。JobMate 则围绕可选择的同伴案例和持续追问重新组织职业探索；对话还促使用户说出此前保持内隐的个人限制和判断标准。特别是，参与者可以通过接受、质疑或修订来回应 agent 对其能力、经历和职业适配性的解释。当 persona 的匹配依据不清、来源语料中缺少用户想了解的职业路径，或者回应仍然较为宽泛时，参与者选择某个 persona 或继续对话的理由就会减弱。

This work makes three contributions to HCI research on AI-supported job seeking and career exploration. First, we present JobMate, a research probe that transforms authentic peer-authored career posts into source-traceable conversational personas and combines structured case selection with SDT-guided, retrieval-augmented dialogue. Second, we provide an empirical account of how changing peer experience from a browsable post into a conversational counterpart reshapes career exploration. Our findings show how largely implicit social comparison can become explicit and open to discussion when users explain their own situations, respond to the agent's interpretations, and reconsider what a peer's experience means for them. Third, we derive design considerations for AI systems built on peer experiences: clearly separate peer-authored content from model interpretations, explain why each persona is recommended, and help users state their constraints and correct the system's interpretations of them.
%% 本研究对 AI 支持的求职和职业探索相关 HCI 研究作出三项贡献。第一，我们提出 JobMate：一个把真实的同伴职业帖子转化为来源可追溯的对话式 persona，并将结构化案例选择与 SDT 引导的检索增强对话结合起来的研究 probe。第二，我们从实证上说明，把同伴经历从可浏览的帖子变成对话对象，会如何改变职业探索。我们的发现表明，当用户说明自己的情况、回应 agent 的解释，并重新思考同伴经历对自己的意义时，原本大多保持内隐的社会比较可以变得明确，也可以在对话中被讨论。第三，我们提出同伴经验型 AI 系统的设计需求：清楚区分同伴原文与模型解释，说明推荐每个 persona 的原因，并帮助用户表达自身限制以及纠正系统对自己的解释。

%% ============================================================
%% 2. RELATED WORK 
%% ============================================================
\section{Related Work}
\label{sec:relatedwork}

\subsection{Peer Career Experiences on Social Media}
\label{subsec:rw-peer}

For young jobseekers with little first-hand work experience, what is missing from official career resources is not a description of what jobs exist, but an account of what happened to people like them afterwards; such knowledge is often conveyed through narratives that ordinary users choose to make public. Recent work shows that career-oriented posts on social platforms expose users to a wider range of trajectories and help them imagine and prepare possible selves~\cite{Corvite2024}, and that young jobseekers now treat social media as a primary channel for job-related information~\cite{Mowbray2021}. The value of these posts is precisely that each is a specific person's specific choice in a specific situation, which is also why they cannot be applied to a reader in the way a job description can.
%% 对于几乎没有一手工作经验的年轻求职者来说，正规的职业资源缺的不是"世界上有哪些岗位"，而是"和我类似的人后来怎么样了"；这类知识几乎只存在于普通用户自愿公开的叙述里。近期研究表明，社交平台上的职业内容让用户看到更多样的职业轨迹，并据此想象和准备 possible selves~\cite{Corvite2024}；年轻求职者也已经把社交媒体当作获取求职信息的主要渠道~\cite{Mowbray2021}。这些帖子的价值恰恰在于它们是"具体的人在具体处境下做出的具体选择"——这也意味着它们无法像职位描述那样被直接套用到读者身上。

Career exploration has long been understood as involving both exploration of the self and exploration of educational and occupational environments~\cite{Stumpf1983CareerExploration,Flum2000VocationalExploration,Jiang2019CareerExploration}. Peer posts bring these two forms together: users learn about an external path while simultaneously considering their own interests, abilities, and constraints.

Reading such posts also has emotional consequences, and what existing research has documented is their negative side: career frustration attributed to social comparison on career-related content \cite{Fukubayashi2021}, job-search pressure from social media \cite{Zhang2024Pressure}, and employment anxiety among Chinese youth exposed to such content \cite{Luo2024}. Social comparison theory offers a partial account of why this happens, noting that the direction of comparison shapes its affective valence and that upward comparison can both motivate and diminish self-evaluation \cite{Festinger1954,Buunk1990,Wills1981,Vogel2014SocialComparison,Appel2023}. Yet these studies measure the \emph{consequences} of comparison, such as anxiety, pressure, or frustration, rather than how comparison actually unfolds during browsing. They leave open a question that the same evidence naturally raises: why the very act of gathering information from these posts tends to produce anxiety alongside it. Existing work thus establishes that peer career posts matter and that their emotional cost is real, but it treats \emph{browsing} as a single independent variable. We therefore do not know what makes users decide whom to read and whom to skip, nor what happens in the interval between reading others' experiences and arriving at frustration or anxiety. We therefore explore this process through a new form of interaction.
%% 阅读这类帖子也是有情绪后果的，而已有研究记录到的是其负面一侧：与职业内容上的社会比较相关联的职业挫败感 \cite{Fukubayashi2021}、社交媒体带来的求职压力 \cite{Zhang2024Pressure}，以及暴露于此类内容的中国年轻人所报告的就业焦虑 \cite{Luo2024}。社会比较理论对为什么会这样给出了部分解释：比较的方向决定其情绪效价，向上比较既可能激励，也可能贬损自我评价 \cite{Festinger1954, Buunk1990, Wills1981, Appel2023}。但这些研究测量的是比较的\emph{后果}——焦虑、压力、挫败——而不是比较在浏览过程中如何实际发生，也留下了同一批证据自然引出的一个问题：为什么从这些帖子里获取信息的过程本身，往往会同时带来焦虑。因此已有工作确立了同伴职业帖子的价值及其情绪代价，但普遍把\emph{浏览}当作单一自变量。我们于是不知道两件事：用户凭什么决定看谁、跳过谁（$\rightarrow$~Theme~1）；以及从阅读他人经历到形成职业挫败感与就业焦虑之间，究竟发生了什么（$\rightarrow$~Theme~4）。因此我们尝试通过新的交互方式来探索这个过程的发生方式。

\subsection{Supporting Sensemaking of User-Generated Experiences}
\label{subsec:rw-sensemaking}

Faced with the fragmentation and noise of community-authored content, HCI has largely responded by having the system take over the work of filtering, organising, and synthesising. They reduce content to what a reader is likely to need: ComViewer supports viewers of online mental-health communities through visual filtering, highlighting, summarisation, and a question panel over posts and comments~\cite{Wu2025ComViewer}. Or restructuring information for exploration rather than shortening it: Sensecape supports complex information-seeking through multi-level abstraction and spatial organisation~\cite{Suh2023Sensecape}, and Graphologue turns linear LLM answers into interactive node-link diagrams that expose the relations inside a topic~\cite{Jiang2023Graphologue}. And extracting task-relevant fragments from unstructured discussion, as PlanHelper pulls actionable points out of question-answering communities to support plan construction~\cite{Li2022PlanHelper}. Career-focused systems likewise structure jobs, skills, and career paths through interactive knowledge exploration, profiling, matching, and visualisation~\cite{Ashok2018JobSense,JoseGarcia2023C3IoC}. These designs share a broad premise: fragmented, noisy content can consume users' limited cognitive resources, motivating systems to support filtering, organisation, and synthesis~\cite{Pirolli2005,Sweller1988}. They help users understand the content.
%% 面对社区内容的碎片化与噪声，HCI 的主流回应是让系统承担筛选、组织与综合的工作。第一条路径是把内容压缩到读者可能需要的部分：ComViewer 通过可视化过滤、高亮、摘要以及针对帖子和评论的提问面板，支持在线心理健康社区中的浏览者~\cite{Wu2025ComViewer}。第二条路径不是缩短内容，而是重组信息结构以便探索：Sensecape 通过多层抽象与空间组织支持复杂信息检索~\cite{Suh2023Sensecape}，Graphologue 把线性的 LLM 回答转化为交互式节点—链接图，使话题内部的关系被显式呈现~\cite{Jiang2023Graphologue}。第三条路径从非结构化讨论中抽取任务相关的片段：PlanHelper 从问答社区中提取可用于计划的要点~\cite{Li2022PlanHelper}。这些设计共同的前提是：碎片与噪声会消耗有限的认知资源，因而值得由系统代为承担~\cite{Pirolli2005, Sweller1988}。它们也共享同一个成功判据：用户理解了内容。

Moving from static reading to conversational search changes what a reader can do with community content, but it also introduces new losses. ConSearcher supports conversational information-seeking in online communities through seeker/provider personas, and in doing so surfaces its own difficulties around persona selection and over-personalisation~\cite{Wu2026ConSearcher}; Sharma et al.~further caution that generative synthesis is not neutral and can narrow the diversity of what users end up seeing~\cite{Sharma2024EchoChamber}. That interactive questioning changes not only what is retrieved but how deeply it is processed is well established at the level of engagement~\cite{Chi2014}. Across filtering, restructuring, extraction, and conversational search, however, what is being optimised is the intelligibility of the \emph{content}. For a jobseeker, understanding a post is not the same as knowing whether it applies to them: the same paragraph can be reassuring or irrelevant depending on where the reader lives, what they can commit to each week, and what they are already qualified for. Existing systems have primarily focused on making content easier to access and understand, leaving less examined how users relate that content to their own circumstances. Does the system supply more knowledge, or the user supply more of their own situation?
%% 从静态阅读转向对话式检索，改变了读者能对社区内容做什么，但也带来了新的损失。ConSearcher 通过 seeker/provider 两类 persona 支持在线社区中的对话式信息检索，同时报告了 persona 选择与过度个性化方面的问题~\cite{Wu2026ConSearcher}；Sharma 等进一步提醒，生成式的综合并非中立，可能收窄用户最终看到的信息多样性~\cite{Sharma2024EchoChamber}。交互式提问改变的不只是检索到什么，也改变加工深度，这在参与层次上已有明确论证~\cite{Chi2014}。然而，无论是过滤、重组、抽取还是对话式检索，被优化的都是\emph{内容}的可理解性。对求职者而言，读懂一条帖子并不等于知道它是否适用于自己：同一段话是让人安心还是与自己无关，取决于读者住在哪里、每周能投入多少、以及自己已经具备哪些资格。现有系统止步于内容变得可读的那一刻；它们较少追问：泛泛的社区信息要在什么条件下才会对某个具体的人变得可执行，以及这一步究竟由谁完成——是系统提供了更多知识，还是用户交代了更多关于自己的处境（$\rightarrow$~Theme~2）。

\subsection{Persona-Grounded Conversational Agents and the Self}
\label{subsec:rw-persona}

When a conversational agent is given a stance, users do not merely extract information from it; they also revisit their own situation through that stance. Prior digital career interventions include mobile self-reflection activities, scheduled guidance messages, and conversational advising for university students~\cite{Soares2022CareerInterventions,AragonHahner2023FindMyself,AragonBartsch2021DayInLife,Lee2021CASExplorer}. Related LLM systems use adaptive questions to keep reflection user-directed~\cite{Song2025ExploreSelf}. In the career domain, two design patterns have emerged. In one, the agent takes on a normative guiding role: career chatbots assess interests or skills, recommend options, or guide learning paths~\cite{DSilva2020CareerChatbot,Nguyen2022ITCareerBot}, while a self-determination-theory-based career-counselling chatbot supports autonomous decision-making and eases decisional difficulty~\cite{Han2025SDT}, and CareerSim combines LLMs and gameful elements to prompt career-related reflection~\cite{Du2024CareerSim}. In the other, the agent is a projection of the user themselves: dialogues with a letter- or voice-mediated future self improve goal clarity and future-self continuity~\cite{Jeon2025LettersFutureSelf,Pataranutaporn2024FutureYou}. Part of what these effects rely on is well understood: articulating one's own situation to another party is itself a psychological and relational event~\cite{Ho2018,Collins1994}, and JobMate's conversational design draws on the same tradition of autonomy, competence, and relatedness~\cite{Ryan2000}. Across both patterns, however, the speaker on the other side is either a normative role or the user themselves---no third party is in the room.
%% 当对话 agent 被赋予某种立场时，用户不只是从中抽取信息，也会通过这个立场重新审视自己的处境。在职业场景中，出现了两种设计模式。第一种把 agent 设定为规范性的引导者：基于自我决定理论的职业咨询 chatbot 支持自主决策并缓解决策困难~\cite{Han2025SDT}，CareerSim 则结合 LLM 与游戏化元素以促成职业反思~\cite{Du2024CareerSim}。第二种把 agent 设定为用户自身的投射：通过与信件或语音形式的未来自我对话，提升目标清晰度与未来自我连续性~\cite{Jeon2025LettersFutureSelf, Pataranutaporn2024FutureYou}。这些效果所依赖的机制有一部分已经被清晰阐明：向另一方说出自身处境本身就是一件心理与关系事件~\cite{Ho2018, Collins1994}，JobMate 的对话设计也承接了 autonomy、competence、relatedness 这一传统~\cite{Ryan2000}。然而，在这两种模式下，坐在对面说话的要么是一个规范角色，要么就是用户自己——没有第三个人在场。

Persona-based agents have brought a third party into the conversation, but the credibility of that third party still comes from its specification rather than from anything a real person has actually lived through. Agents with divergent stances expose users to a wider range of viewpoints on contested issues~\cite{Zhang2024SeeWidely}; simulated expert personas support research ideation and perspective-taking~\cite{Liu2025PersonaFlow}; generative agents have been shown to sustain believable behaviour across extended interactions~\cite{Park2023GenerativeAgents}; and CloChat examines how users customise and inhabit LLM personas~\cite{Ha2024CloChat}. However, the personas above are fictional experts, generated stances, or projected future selves, whereas each persona in JobMate is anchored to a specific career narrative that a real peer has publicly written, with the original post kept side-by-side so that any claim can be checked against it. What's more, those personas mainly project outward, offering views on a topic; a persona anchored to a real peer's experience additionally speaks back to the user---about what their own trajectory is worth, what they might be suited for, what they are missing. The user therefore has to deal with an external interpretation of themselves rather than only receive information. It is precisely this reverse direction that makes acceptance, contestation, and revision possible; browsing peer posts has no counterpart for it, because there is no specific speaker to answer back to. We accordingly know little about how people engage a conversational agent grounded in specific peer-authored career experiences---particularly how, when such an agent offers an interpretation of the user themselves, that interpretation becomes something users can accept, contest, or revise, and how this differs from the private, unaddressable comparison that ordinary browsing affords. We approach this question by turning real peer posts into conversational personas and using the resulting system as a research probe alongside free browsing on the same platform, so that the transformation itself can be observed rather than assumed.
%% 基于 persona 的 agent 已经把第三方带进对话，但这一第三方的可信度仍然来自其设定，而不是来自某个真实的人真正经历过什么。持不同立场的 agent 让用户接触到关于争议话题的更多样观点~\cite{Zhang2024SeeWidely}；模拟专家 persona 支持科研构思与视角切换~\cite{Liu2025PersonaFlow}；生成式 agent 已被证明可以在长时交互中维持可信的行为~\cite{Park2023GenerativeAgents}；CloChat 则考察用户如何定制并体验 LLM persona~\cite{Ha2024CloChat}。有两点把这条线索与我们研究的对象区分开来。\emph{来源}：上述 persona 是虚构专家、生成立场或未来自我的投射，而 JobMate 中的每个 persona 都锚定在一位真实同伴公开写下的具体求职经历上，原帖始终并排可见，使任何断言都能被回查。\emph{方向}：这些 persona 主要向外投射观点，就某个话题给出看法；而锚定在真实同伴经历上的 persona 会额外\emph{反过来}对用户本人说话——你这段轨迹值多少、你可能适合什么、你缺什么。用户因此必须处理一个外部主体对自己的解释，而不只是接收信息。正是这个反向的方向，使接受、反驳与修订成为可能（$\rightarrow$~Theme~3）；浏览同伴帖子时没有与之对应的东西，因为没有一个具体的说话人可以被回应。因此我们仍然所知甚少：当对话 agent 锚定在具体同伴写下的求职经历上时，人们如何与之互动——尤其是当这样的 agent 主动给出对用户本人的解释时，这种解释如何成为可以被接受、反驳或修订的东西，以及它与普通浏览所提供的那种无处回应的、私下的比较又有何不同。我们通过把真实的同伴帖子转化为可对话的 persona，并将由此形成的系统作为一件研究 probe，与同一平台上的自由浏览并置，从而使这一转化本身可以被观察，而不是被预设。

\section{Method}
\subsection{Overview}

Our study examines how transforming peer career experiences from browsable social-media posts into persona-grounded conversational encounters reshapes how job seekers make sense of those experiences in relation to themselves. We pursued this question in three phases. First, a formative study (Section~\ref{sec:formative}) established job seekers' existing practices and the central tension of peer-post browsing: specific accounts provide valuable references, but finding and applying them requires substantial cognitive and emotional work. We then built JobMate (Section~\ref{sec:jobmate}) as a research probe that enacts this transformation by turning static posts into source-visible conversational personas, making users' questions, interpretations, and responses observable. Finally, a between-subjects study with 24 participants (Section~\ref{sec:userstudy}) compared JobMate with native RedNote browsing as two analytic views of the same sensemaking process, rather than as a treatment and a control. Quantitative measures characterize trends in this small sample, while the primary analysis draws on interviews and interaction logs.
%% 本研究考察：将同伴职业经历从可浏览的社交媒体帖子转化为基于 persona 的对话式相遇，会如何改变求职者联系自身来理解这些经历的方式。我们通过三个阶段研究这一问题。首先，形成性研究（第~\ref{sec:formative}~节）确认了求职者现有的实践，以及浏览同伴帖子时的核心张力：具体经历能够提供有价值的参照，但寻找这些经历并判断其如何适用于自身，需要投入大量认知和情绪劳动。随后，我们构建了 JobMate（第~\ref{sec:jobmate}~节），将其作为一件研究 probe，通过把静态帖子转化为来源可见的对话式 personas 来实现这一转变，使用户的问题、解释和回应成为可观察的材料。最后，我们开展了一项包含 24 名参与者的组间研究（第~\ref{sec:userstudy}~节），将 JobMate 和原生小红书浏览作为观察同一意义建构过程的两个分析视角，而不是一组实验处理与控制条件。定量测量用于呈现这一小样本中的趋势，主要分析则建立在访谈和交互日志之上。

\subsection{Formative Study}
\label{sec:formative}

\subsubsection{Procedure}

To understand job seekers' existing practices, pain points, and needs when browsing peer-authored career posts, we conducted a formative study combining an online survey with follow-up semi-structured interviews.
%% 为了理解求职者浏览同伴职业帖子时的现有实践、痛点和需求，我们开展了一项结合在线问卷与后续半结构化访谈的形成性研究。

We distributed the survey at a local university and received 64 valid responses from students who had used social media to seek career information (45 women and 19 men; ages 19--33, $M=24.1$, $SD=3.1$). The survey combined multiple-choice items with open-ended questions about respondents' job-search difficulties, use of career-related posts, problems encountered during browsing, and expectations for AI assistance. We then interviewed eight participants from eight disciplines (F1--F8; 6 women and 2 men; ages 22--28, $M=24.3$). Each interview lasted 30--60 minutes and examined participants' information-seeking practices, emotional experiences, attempts to seek advice from others, and expectations for career-support tools. All interviews were audio-recorded and transcribed. We used the survey to identify broad concerns and analyzed the interview transcripts to identify recurring practices, difficulties, and needs.
%% 我们在当地一所大学发放问卷，从曾使用社交媒体寻找职业信息的学生中获得了 64 份有效回答（45 女、19 男；年龄 19--33 岁，$M=24.1$，$SD=3.1$）。问卷结合选择题与开放式问题，询问参与者的求职困难、职业类帖子的使用方式、浏览中遇到的问题，以及对 AI 辅助的期待。随后，我们访谈了来自八个专业的八名参与者（F1--F8；6 女、2 男；年龄 22--28 岁，$M=24.3$）。每次访谈持续 30--60 分钟，重点了解参与者获取信息的方式、情绪体验、向他人寻求建议的经历，以及对职业支持工具的期待。所有访谈均被录音并转录。我们通过问卷识别普遍问题，并分析访谈转录文本，以归纳反复出现的实践、困难和需求。

\subsubsection{Findings}

\textbf{Participants used peer-authored posts to access concrete experiences unavailable in formal job descriptions or their immediate social circles.} They searched for interview processes, salaries, offer outcomes, workplace evaluations, and preparation timelines, and used peers with similar backgrounds to judge whether a path was attainable. These accounts also broadened the set of paths participants could imagine. F8, for example, explained that RedNote showed students of traditional Chinese medicine options beyond hospitals, ranging from civilian positions in the military to pet acupuncture. Peer posts therefore served both as practical guidance and as examples of careers participants had not previously considered.
%% 参与者通过同伴帖子获得正式职位描述或身边社交圈难以提供的具体经历。他们搜索面试流程、薪资、offer 结果、工作评价和准备时间线，并通过背景相似的同伴判断某条路径是否可行。这些经历也拓展了参与者能够想象的职业范围。例如，F8 表示，小红书让中医学生看到医院之外的选择，从军队文职到宠物针灸。因此，同伴帖子既提供了实际指导，也让参与者认识到此前未曾考虑的职业可能性。

\textbf{Finding useful cases required participants to filter fragmented and uneven information.} Advertisements and paid-service promotions appeared alongside personal accounts (F1, F4, F6, F8), while niche disciplines sometimes had few relevant posts (F4). Participants also encountered conflicting evaluations and had to reconstruct whether an author's background resembled their own before using the experience as a reference. F7 described having to judge background fit while reading, which made the process feel like \emph{``cognitive overload.''} The value of peer experience therefore depended on substantial work to locate, verify, and compare suitable cases.
%% 然而，找到有用案例需要参与者筛选碎片化且质量不一的信息。广告和付费服务推广常与个人经历混在一起（F1、F4、F6、F8），小众专业有时又缺少相关帖子（F4）。参与者还会遇到彼此冲突的评价，并且必须先还原发帖者的背景是否与自己相似，才能判断其经历能否作为参照。F7 将这种边读边判断背景匹配的过程形容为“认知过载”。因此，同伴经历的价值依赖于用户投入大量精力寻找、核实和比较合适的案例。

\textbf{Browsing peer experiences could intensify anxiety because participants had limited control over the comparisons presented to them.} Reports of industry decline, highly successful job seekers with multiple offers, and exaggerated negative accounts all became grounds for evaluating one's own prospects. F1 felt anxious both about a declining industry and about others receiving many offers, while F5 described exaggerated claims as producing \emph{``catastrophic imagination''} about the future. F7 likewise found that conflicting accounts made her more anxious as she continued reading. Social media thus supplied useful reference points while repeatedly exposing users to comparisons that were difficult to interpret or disengage from.
%% 浏览同伴经历也可能加剧焦虑，因为参与者难以控制平台向他们呈现怎样的比较对象。行业衰退、他人获得多个 offer，以及被夸大的负面经历，都可能成为他们评价自身前景的依据。F1 既会因行业低迷而焦虑，也会因他人获得多个 offer 而焦虑；F5 则把夸张信息造成的影响描述为对未来的“灾难化想象”。F7 同样发现，互相冲突的评价使她越看越焦虑。因此，社交媒体虽然提供了有用的参照，也会不断把用户置于难以解释或摆脱的比较之中。

\textbf{Static posts rarely resolved questions specific to an individual's situation.} Participants tried posting their own questions, commenting, or contacting authors, but these attempts did not reliably produce a response. F3 reported that private messages often went unanswered, and F5 noted that unclear details in a post could not be clarified immediately. Participants consequently wanted support that could respond to their own background, including help with self-assessment, career-path comparison, resumes, and interview preparation. These findings motivated JobMate's focus on preserving authentic peer accounts while structuring cases for comparison and enabling continued questions through conversation.
%% 静态帖子通常无法解决针对个人处境的具体问题。参与者会尝试自己发帖、评论或联系原作者，但这些方式并不能稳定地得到回应。F3 表示私信经常无人回复，F5 则指出，帖子中不清楚的细节无法被即时澄清。因此，参与者希望获得能够回应自身背景的支持，包括自我评估、职业路径比较、简历修改和面试准备。这些发现促使 JobMate 在保留真实同伴经历的同时，对案例进行结构化以支持比较，并通过对话让用户能够继续追问。

%% ============================================================
%% 4. JOBMATE
%% ============================================================
\begin{figure*}[h]
  \centering
  \includegraphics[width=0.95\textwidth]{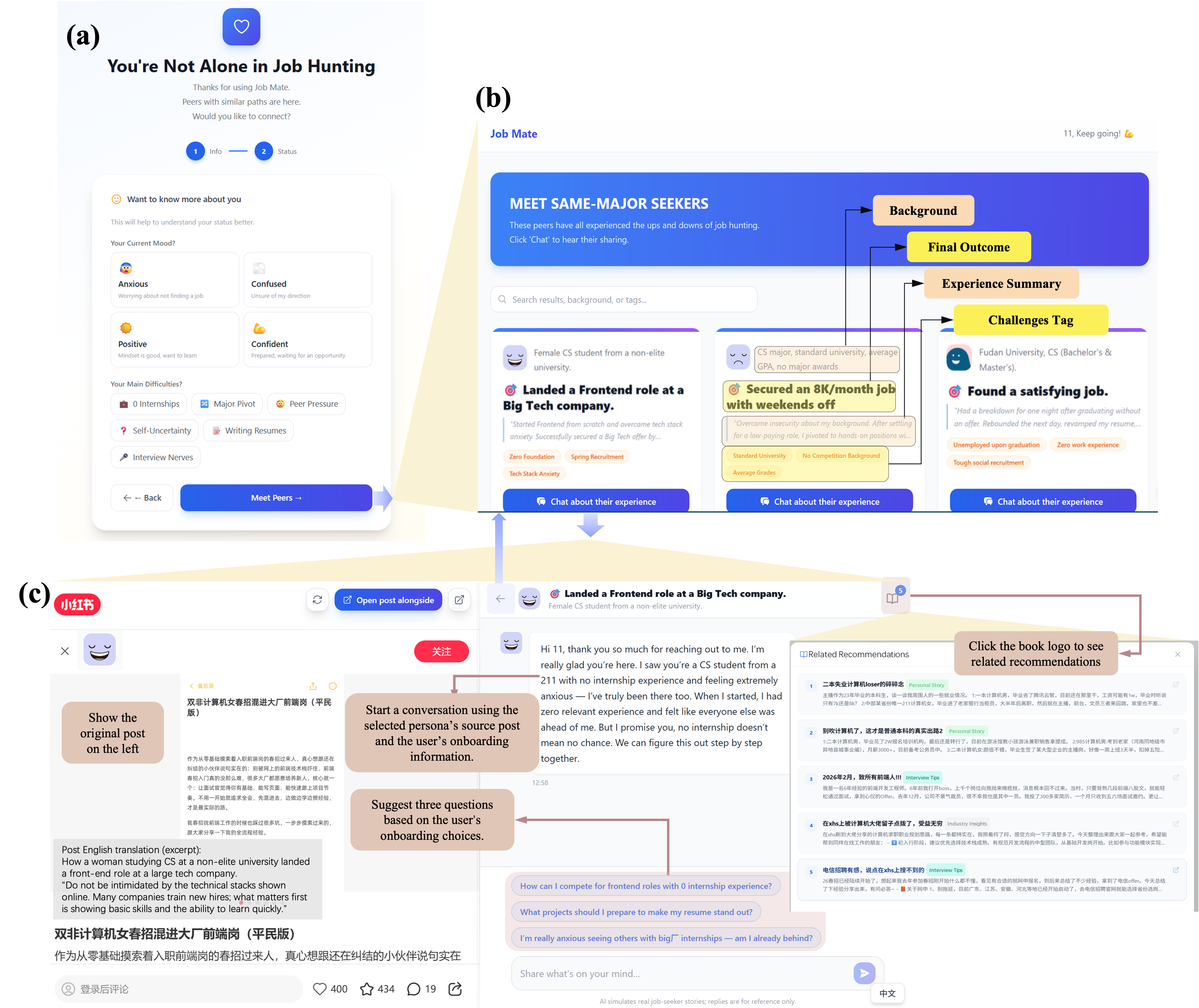}
  \caption{JobMate interfaces from onboarding to persona-grounded exploration. Users provide their background and current concerns, browse persona cards that summarise each peer through four consistent fields, and explore a chosen persona through a conversation shown alongside the original post. Suggested questions and related posts provide additional ways to continue exploring.}
  \Description{A composite interface figure showing the JobMate user journey from onboarding to persona cards and a source-visible conversation with suggested questions and related posts.}
  \label{fig:interface}
\end{figure*}

\begin{figure*}[h]
  \centering
  \includegraphics[width=0.90\textwidth]{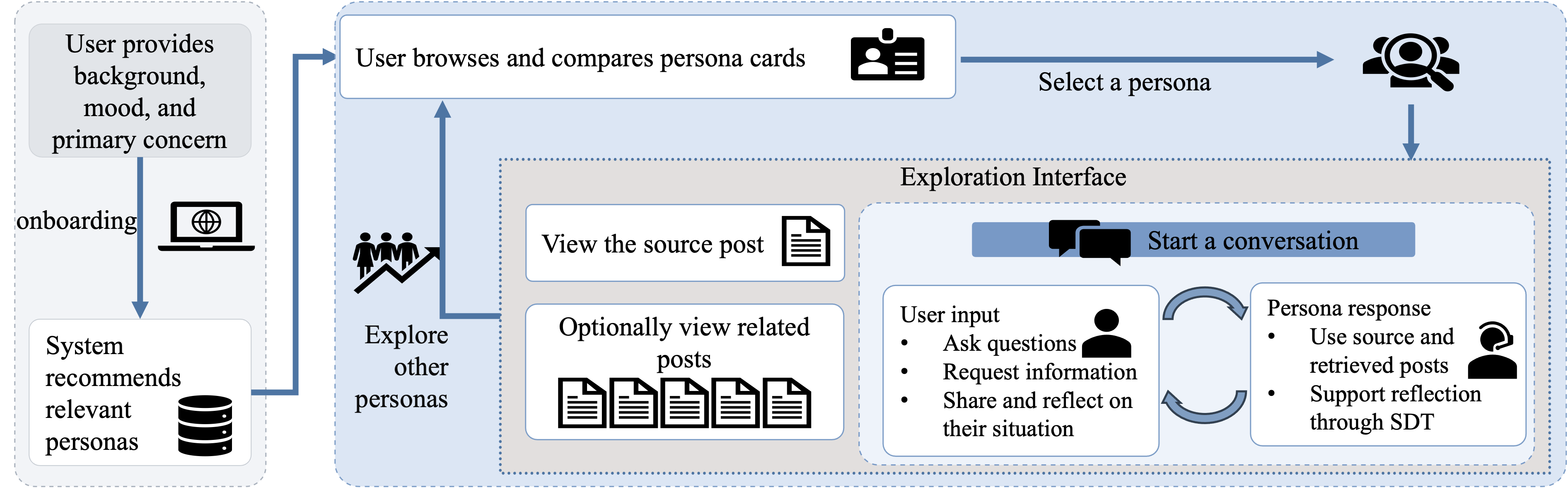}
  \caption{JobMate interaction flow from onboarding and browsing persona cards to source-visible conversation, optional related reading, and exploration of other personas.}
  \Description{A workflow showing users providing background and current concerns, browsing persona cards, opening a source post and conversation, viewing related posts, and returning to explore another persona.}
  \label{fig:interaction}
\end{figure*}
\clearpage

\subsection{JobMate as a Research Probe}
\label{sec:jobmate}

JobMate transforms career posts written by real peers into persona-grounded conversational agents. We use it as a research probe~\cite{Gaver1999, Boehner2007}: in addition to supporting career exploration, it makes visible how users select peer cases, pursue missing information, and relate others' experiences to themselves. We first describe the user journey through JobMate, then explain how posts were collected and organised into personas, and finally detail the post-grounded conversation mechanism.
%% JobMate 将真实同伴撰写的职业帖子转化为基于 persona 的对话 agent。我们将其作为一个 research probe~\cite{Gaver1999, Boehner2007}：除了支持职业探索之外，它还使用户如何选择同伴案例、追问缺失信息，以及把他人的经历关联到自身变得可见。下面我们首先介绍用户如何使用 JobMate，随后说明帖子如何被收集、整理并构建为 personas，最后介绍基于原帖的对话机制。

\subsubsection{Using JobMate}
\label{sec:using-jobmate}

Figure~\ref{fig:interface} presents the interfaces users encounter throughout JobMate. During onboarding, users provide their educational background, current mood, and primary career difficulty. JobMate then recommends a gallery of personas from the same discipline. Each card summarises a peer through four consistent fields---background, job-seeking outcome, experience summary, and challenges---so that users can browse and learn about the available cases before opening a full post. Selecting \emph{Chat about their experience} opens a split-view exploration interface: the original RedNote post remains visible on the left, while the selected persona and three starter questions tailored to the onboarding information appear on the right. Users can ask their own questions, disclose further context, or select a starter question. They can also open a panel containing five related posts or return to the gallery and choose another persona.
%% 图~\ref{fig:interface}展示了用户在 JobMate 中依次接触到的界面。在 onboarding 阶段，用户提供教育背景、当前情绪和主要职业困难。随后，JobMate 推荐一组来自相同学科的 personas。每张卡片采用一致的四个字段——背景、求职结果、经历摘要和困难——概括一位同伴，使用户能够在打开完整帖子前浏览并了解可选案例。点击 \emph{Chat about their experience} 后，用户进入分栏式探索界面：小红书原帖持续显示在左侧，所选 persona 与三个根据 onboarding 信息生成的起始问题显示在右侧。用户可以自行提问、补充个人情境，或直接选择一个起始问题；也可以打开包含五篇相关阅读的面板，或返回卡片列表选择另一位 persona。

Figure~\ref{fig:interaction} abstracts these screens into the complete interaction flow. The flow is deliberately iterative rather than terminal: within a selected case, users can move among the source post, related posts, and conversation; when a case no longer addresses their concern, they can return to the gallery and explore another persona. The probe therefore captures both how users choose among peers and how their questions and self-descriptions develop within a chosen case.
%% 图~\ref{fig:interaction}将上述界面概括为完整的交互流程。该流程被设计为可反复进行，而不是以单一终点结束：在选定的案例中，用户可以在原帖、相关阅读和对话之间移动；当当前案例不再回应其问题时，也可以返回卡片列表探索另一位 persona。因此，这一 probe 既能呈现用户如何在不同同伴之间作出选择，也能记录其问题和自我描述如何在选定案例的对话中逐步发展。

\subsubsection{Post Collection and Persona Construction}
\label{sec:pipeline}

\begin{figure*}[t]
  \centering
  \includegraphics[width=0.85\textwidth]{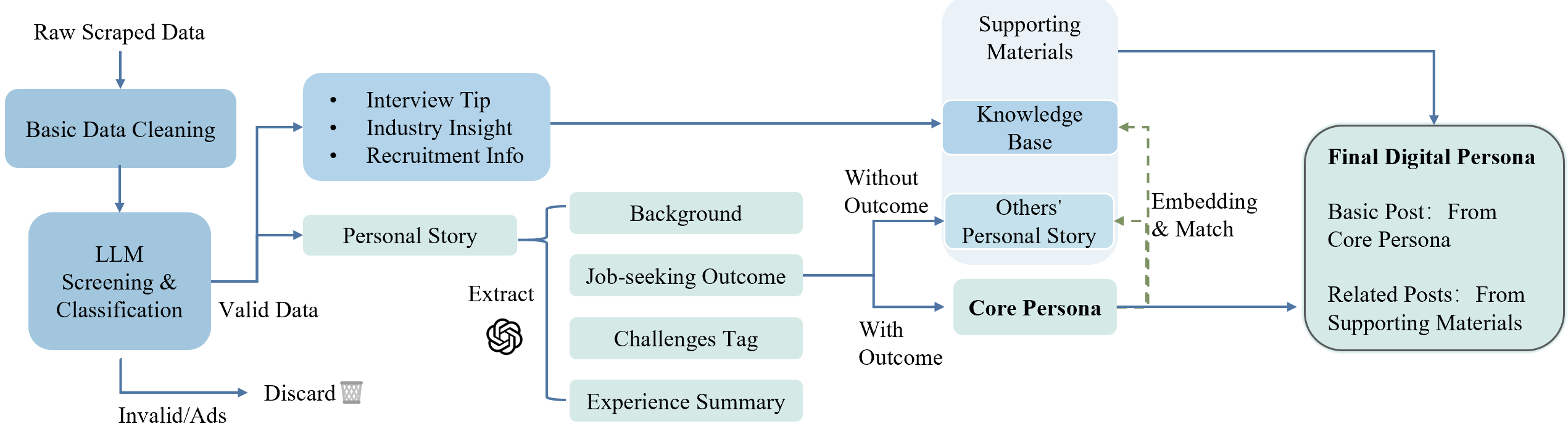}
  \caption{How JobMate turns RedNote posts into conversational personas. After cleaning and classifying the collected posts, the system extracts four fields from personal career stories. Stories with a complete career trajectory become personas, while the remaining personal stories and practical posts are retrieved to support their conversations.}
  \Description{A pipeline diagram showing four stages. First, raw social posts are cleaned and classified. Second, valid posts are split into personal stories and a knowledge base. Third, stories are structured into background, outcomes, challenges, and summary, with complete narratives entering a core persona pool. Fourth, each final digital persona is assembled from one core post plus retrieved related stories and practical resources.}
  \label{fig:pipeline}
\end{figure*}

Figure~\ref{fig:pipeline} shows how RedNote posts were collected and organised into the materials used by JobMate. We collected 660 career-related posts across Computer Science, Psychology, and Chinese Literature. Deduplication, text cleaning, and length filtering retained 414 posts; automated validity filtering and classification then retained 364 posts and organised them as personal experiences, interview tips, industry insights, or recruitment information. From personal career narratives, we extracted four fields for consistent comparison: background, job-seeking outcome, an experience summary, and one to three challenges. The 118 posts with a complete career trajectory became the core persona pool, with each persona anchored to one identifiable source post rather than a synthetic biography. The remaining materials formed two retrieval resources: related peer experiences and practical information. This organisation preserves source traceability while also retaining gaps in the platform's coverage. Appendix~\ref{app:pipeline} reports the models, verification procedure, and complete corpus statistics.
%% 图~\ref{fig:pipeline}展示了小红书帖子如何被收集并整理为 JobMate 使用的材料。我们从计算机、心理学和中文三个学科共收集了 660 篇职业相关帖子。经过去重、文本清洗和长度筛选后保留 414 篇；随后通过自动有效性筛选和分类保留 364 篇，并将其整理为个人经历、面试经验、行业见解或招聘信息。对于个人职业叙事，我们抽取四个字段以支持一致比较：背景、求职结果、经历摘要和一至三个困难标签。具有完整职业轨迹的 118 篇帖子构成核心 persona 池；每个 persona 都锚定在一篇可识别的原帖上，而不是一份合成的人物传记。其余材料则形成两类检索资源：相关同伴经历和实用信息。这种组织方式在保留来源可追溯性的同时，也保留了平台语料自身的覆盖缺口。附录~\ref{app:pipeline}报告所用模型、校验过程和完整语料统计。

\subsubsection{Post-Grounded Conversation Mechanism}
\label{sec:sdt}
\label{sec:conversation}

For each selected persona, the conversation prompt combines five inputs: the persona's source post, its four structured fields, the user's onboarding information, retrieved materials, and conversational rules. Retrieval follows two tracks. The first supplies two peer experiences matched by background or difficulty; the second supplies three practical resources related to the persona's career destination. Rather than speaking on behalf of the original author, the system presents an AI-generated persona grounded in the peer's published account. The continuously visible source allows users to inspect the experiential basis of its responses and distinguish the original account from generated dialogue.
%% 对于每个被选中的 persona，对话 prompt 结合五类输入：该 persona 的原帖、四个结构化字段、用户的 onboarding 信息、检索材料和对话规则。检索分为两条路径：第一条提供两篇在背景或困难方面相似的同伴经历；第二条提供三篇与该 persona 职业去向相关的实用信息。系统并不代表原作者发言，而是呈现一个锚定于同伴公开叙述的 AI 生成 persona。持续可见的原帖使用户能够检查回答所依据的真实经历，并区分原始叙述与生成式对话。

The conversational rules operationalise Self-Determination Theory~\cite{Ryan2000}. \textbf{Relatedness} connects difficulties documented in the peer's post to the user's situation; \textbf{competence} helps users recognise abilities in experiences they may undervalue; and \textbf{autonomy} presents options rather than directives and leaves room for disagreement or withdrawal. The mechanism is designed to elicit articulation rather than agreement. By asking questions and supplying personal context, users make previously implicit constraints and criteria explicit; when the agent interprets their abilities, experiences, or career fit, they can accept, challenge, or revise that interpretation. These multi-turn exchanges make the sensemaking process observable while keeping the source available for verification. Complete prompting and implementation details are reported in Appendix~\ref{app:implementation}.
%% 对话规则对自我决定理论~\cite{Ryan2000}进行了操作化：\textbf{关系需求}将原帖中记录的同伴困难与用户处境联系起来；\textbf{胜任需求}帮助用户从可能被自己低估的经历中识别能力；\textbf{自主需求}提供选项而不是指令，并为不同意或退出保留空间。该机制旨在引出用户的表达，而不是要求用户认同 agent。通过提问和补充个人情境，用户会将原本内隐的限制和判断标准明确表达出来；当 agent 解释其能力、经历或职业适配性时，用户则可以接受、质疑或修订这种解释。这些多轮交流使意义建构过程变得可观察，同时保留原帖以供核查。完整 prompt 和实现细节见附录~\ref{app:implementation}。

\subsection{Comparative Study}
\label{sec:userstudy}

To understand how post-grounded conversational agents shape job seekers’ sense-making of peer career experiences, we conducted a between-subjects study with 24 participants, comparing JobMate with native RedNote browsing. We used this comparison primarily as an analytic lens rather than as an evaluation of which interface performed better. Both conditions involved peer-authored career posts originating from RedNote: native browsing showed how participants ordinarily encountered and interpreted such experiences, whereas JobMate externalised parts of this sense-making process through inspectable conversations. The two conditions therefore provided different views of the same career-exploration process.

Each session produced post-task interview data and questionnaire responses concerning decision difficulty, cognitive load, perceived support, and self-determination. The JobMate condition additionally produced complete conversation logs. Together, these data allowed us to examine both how participants described their reasoning retrospectively and how their judgments developed during interaction.
%%为了理解基于真实同伴帖子的对话式智能体如何影响求职者理解同伴职业经历，我们开展了一项包含24名参与者的组间研究，将JobMate与原生小红书浏览进行比较。该比较主要作为一种分析视角，而不是为了评价哪一种界面表现更好。两个条件都涉及来自小红书的同伴职业帖子：原生浏览呈现参与者通常如何接触和理解这些经历，而JobMate则通过可供分析的对话，将这一意义建构过程的部分内容外显出来。因此，两个条件为同一个职业探索过程提供了不同的观察视角。
%%每场实验都产生了任务后访谈数据，以及关于职业决策困难、认知负荷、感知支持和自我决定的问卷数据。JobMate条件还产生了完整的对话日志。这些数据使我们既能考察参与者事后如何描述自己的思考，也能分析其判断如何在交互过程中逐步形成。

\subsubsection{Participants}

We recruited 24 university students actively preparing for job searches (17 women, 7 men; ages 18--25): eight each from Computer Science, Psychology, and Chinese Literature. Within each discipline, participants were randomly assigned to JobMate ($n{=}4$) or RedNote ($n{=}4$), producing 12 participants in each condition.
%% 我们招募了 24 名正在积极准备求职的大学生（17 名女性、7 名男性；年龄 18--25 岁），计算机、心理学和中文专业各 8 人。在每个专业内部，参与者被随机分配到 JobMate（$n{=}4$）或小红书（$n{=}4$），最终每个条件各有 12 名参与者。

\subsubsection{Study Design and Procedure}

Sessions were conducted remotely by video conference and lasted 50--60 minutes. After an introduction and consent, participants completed the pre-task questionnaire and a 30-minute career exploration task. Both groups were asked to examine the experiences and outcomes of peers with similar backgrounds, seek information that might clarify their direction, and consider possible next steps. JobMate participants received a brief interface walkthrough before using the system freely; RedNote participants searched and browsed discipline-related career posts as they normally would.

Participants then completed the post-task questionnaires and a 10--15 minute semi-structured interview. A shared set of questions asked whether and how the exploration experience had informed participants' career thinking, what informational or emotional support they had obtained, how much cognitive effort it required, and how it compared with their usual use of RedNote. We prompted participants to describe concrete episodes, including information that clarified or complicated a possible direction and moments that made them feel reassured, encouraged, or more confident. In the JobMate condition, follow-up questions examined how the persona cards and their background, outcome, and challenge fields shaped case selection; how the visible source post and related recommendations supported interpretation or checking; and how the conversation supported questioning and reflection. 
%% 实验通过视频会议远程进行，每场持续 50--60 分钟。在研究介绍和知情同意后，参与者完成前测问卷和一项 30 分钟的职业探索任务。两组都需要了解背景相似同伴的经历与去向，寻找可能帮助澄清方向的信息，并思考可能的下一步。JobMate 组先接受简短的界面演示，随后自由使用系统；小红书组则按照日常方式搜索和浏览与本专业相关的职业帖子。

%% 任务结束后，参与者完成后测问卷和一次 10--15 分钟的半结构化访谈。共同的核心问题包括：这次探索是否以及如何影响参与者对职业的思考，他们获得了哪些信息支持或情绪支持，整个过程需要投入多少认知努力，以及这种体验与日常使用小红书有何不同。我们要求参与者描述具体事例，包括哪些信息使某个职业方向变得更清晰或更复杂，以及他们在哪些时刻感到安慰、鼓励或信心增强。对于 JobMate 组，追问进一步考察 persona 卡片及其背景、去向和困难字段如何影响案例选择，可见的原帖和相关推荐如何支持理解或核查，以及对话如何支持提问和反思。

\subsubsection{Measures}

We measured career decision-making difficulty before and after exploration with the 34-item Career Decision-Making Difficulties Questionnaire (CDDQ)~\cite{Gati1996CDDQ}. Items used a 7-point scale, with higher scores indicating greater difficulty. Post-task cognitive load was assessed with the NASA Task Load Index (NASA-TLX)~\cite{Hart1988}, with higher overall scores indicating greater perceived workload. Participants also rated informational and emotional support on two 7-point single items, with higher scores indicating greater support. An adapted 12-item measure of basic psychological need satisfaction~\cite{Standage2005,Xi2019} assessed autonomy, competence, and relatedness on 7-point scales, with higher scores indicating greater satisfaction of these needs. Interviews were audio-recorded and transcribed, and complete conversation logs were collected in the JobMate condition.
%% 我们使用 34 题的职业决策困难问卷（Career Decision-Making Difficulties Questionnaire，CDDQ）~\cite{Gati1996CDDQ}测量职业探索前后的决策困难。题项采用 7 点计分，分数越高表示职业决策困难越大。任务后的认知负荷采用 NASA 任务负荷指数（NASA Task Load Index，NASA-TLX）~\cite{Hart1988}进行评估，总分越高表示参与者感受到的任务负荷越大。参与者还使用两个 7 点单题分别评价信息支持和情绪支持，分数越高表示获得的支持越多。一项经调整的 12 题基本心理需求满足度量表~\cite{Standage2005,Xi2019}使用 7 点计分评估自主、胜任和关系需求，分数越高表示相应心理需求得到的满足程度越高。所有访谈均被录音和转录，JobMate 条件还记录了完整对话日志。

\subsection{Data Analysis}

\textbf{Quantitative analysis.} We first calculated descriptive statistics for all questionnaire measures. Shapiro--Wilk tests were used to examine normality. We then used independent-samples $t$-tests for comparisons between the JobMate and RedNote conditions and paired $t$-tests for pre--post comparisons. The quantitative results were used as contextual evidence for interpreting the qualitative findings rather than as a standalone evaluation of system effectiveness.

%% \textbf{定量分析。}我们首先计算了所有问卷指标的描述性统计结果，并使用 Shapiro--Wilk 检验考察数据的正态性。随后，我们使用独立样本 $t$ 检验比较 JobMate 与小红书条件，并使用配对样本 $t$ 检验分析前后测变化。定量结果被用作解释质性发现的背景证据，而不是对系统效果进行独立评价。

\textbf{Interview analysis.} We conducted a thematic analysis of the post-task interviews using open and axial coding. We first read the transcripts to become familiar with the data. During open coding, relevant segments were assigned short, data-near codes describing participants' navigation strategies, information gained, self-reflection, and emotional experiences. During axial coding, related codes were grouped into broader categories, and we examined the conditions, actions, and consequences connecting them. We then developed candidate themes from these categories and repeatedly compared them with the transcripts. Themes were revised when their boundaries overlapped or when the supporting evidence did not adequately represent the variation across participants and conditions. Representative quotations were selected to illustrate both recurring patterns and meaningful differences between participants.

%% \textbf{访谈分析。}我们采用开放编码和主轴编码对任务后的访谈进行了主题分析。首先，我们通读访谈转录文本以熟悉数据。在开放编码阶段，我们为相关文本片段赋予贴近原始数据的简短 codes，用于描述参与者的浏览策略、获得的信息、自我反思和情绪体验。在主轴编码阶段，我们把相关 codes 归入更广泛的 categories，并考察它们之间的发生条件、参与者行动和结果。随后，我们根据这些 categories 形成候选 themes，并反复将其与访谈文本进行比较。当不同主题的边界重叠，或现有证据无法充分反映参与者与实验条件之间的差异时，我们会进一步修订主题。最后，我们选择具有代表性的引语，用于呈现反复出现的模式以及参与者之间具有意义的差异。

\textbf{Conversation-log and case-level analysis.} We organised the JobMate conversation logs by participant, persona, and conversational turn, and classified relevant exchanges according to the developing themes. Rather than analysing isolated messages, we examined multi-turn sequences to identify how users introduced personal context, how the agent interpreted it, and whether users accepted, challenged, or revised that interpretation. We also conducted case-level analyses of participants whose conversations clearly represented recurring or contrasting interaction patterns. Finally, we compared the log-based observations with participants' interview accounts to examine how reported experiences corresponded with their actions during the interaction.

%% \textbf{对话日志与个案层面分析。}我们按照参与者、persona 和对话轮次整理 JobMate 的对话日志，并依据逐渐形成的主题对相关交流进行归类。分析并不把单条消息孤立看待，而是考察连续的多轮对话，以识别用户如何补充个人情境、agent 如何解释这些信息，以及用户是否接受、反驳或修订这些解释。我们还对能够清楚体现重复模式或对比模式的参与者进行了个案层面分析。最后，我们将日志中观察到的互动过程与参与者的访谈表述进行比较，以考察参与者报告的体验与其实际互动行为如何对应。

\subsection{Ethical Considerations}
\label{sec:ethics}

This study required care for both participants and the authors of public posts. Participants provided informed consent, and we refer to them only through identifiers such as P1--P24 and report only excerpts needed for the analysis. Because public availability does not imply consent for AI simulation, JobMate presents each persona as an AI-generated representation grounded in a post, not as the original author. The interface keeps the source post visible and labels replies as AI-simulated. JobMate may also overstate users' abilities or career fit, so we positioned it as a research probe rather than a career counsellor. It offers options, allows users to disagree, correct the system, or withdraw, and states that replies are for reference. We do not treat its interpretations as objective assessments. These measures do not resolve concerns about author consent, attribution, privacy, or contextual integrity, which we revisit in Section~\ref{sec:limitations}.
%% 本研究同时关注参与者和公开帖子作者的权益。参与者在实验前提供知情同意；我们仅以 P1--P24 等编号指代参与者，并只报告分析所需的片段。公开可见不等于作者同意 AI 模拟其身份，因此 JobMate 将每个 persona 标示为基于帖子的 AI 生成再现，而不是原作者本人。界面持续展示原帖，并标注回答由 AI 模拟。JobMate 对用户能力或职业适配性的解释也可能夸大证据，因此我们将其定位为 research probe，而不是职业咨询工具。系统提供选项，允许用户不同意、纠正系统或退出，并说明回答仅供参考。我们不把系统的解释视为客观评价。这些措施仍不能解决作者同意、署名、隐私和情境完整性问题，相关限制见第~\ref{sec:limitations}节。

%% ============================================================
%% 6. RESULTS
%% ============================================================
\section{Findings}
\label{sec:results}

Questionnaire results provided descriptive context for these qualitative findings (Table~\ref{tab:quant-overview}). Mean career decision-making difficulty decreased in both conditions, with the largest decreases appearing in lack of information. Mean ratings of informational support, emotional support, and self-determination were close across conditions, while mean NASA-TLX effort was lower in JobMate. 
%% 问卷结果为下文的质性发现提供了描述性背景（表~\ref{tab:quant-overview}）。两种条件下的平均职业决策困难均有所下降，其中信息缺乏维度的下降最明显。两组的信息支持、情绪支持和自我决定平均评分较为接近，而 JobMate 组的 NASA-TLX 平均努力程度较低。
\begin{table}[h]
  \caption{Descriptive questionnaire results ($n{=}12$ per condition). Values are $M$ ($SD$), except rows reporting the mean decrease ($\Delta M$).}
  \label{tab:quant-overview}
  \small
  \begin{tabular}{@{}p{0.34\columnwidth}p{0.28\columnwidth}p{0.28\columnwidth}@{}}
    \toprule
    Measure & JobMate & RedNote \\
    \midrule
    CDDQ total, pre & $4.44$ ($0.41$) & $4.01$ ($1.00$) \\
    CDDQ total, post & $4.08$ ($0.46$) & $3.54$ ($0.90$) \\
    CDDQ total, $\Delta M$ & $0.36$ & $0.48$ \\
    CDDQ lack of information, $\Delta M$ & $0.74$ & $1.04$ \\
    NASA-TLX effort & $3.92$ ($1.00$) & $5.08$ ($1.08$) \\
    Informational support & $5.67$ ($1.15$) & $5.58$ ($1.00$) \\
    Emotional support & $5.33$ ($1.37$) & $5.25$ ($1.48$) \\
    Self-determination & $4.66$ ($0.80$) & $4.78$ ($0.56$) \\
    \bottomrule
  \end{tabular}
\end{table}

We next present the qualitative findings. Peer career posts gave participants access to paths beyond their immediate surroundings, but participants still had to decide which cases were relevant and what those experiences meant for them. Turning posts into conversational personas changed how participants navigated cases, developed questions, and related peer experiences to themselves. Sections~\ref{sec:theme1} and~\ref{sec:theme2} describe how participants selected peer cases and obtained career information from them. Section~\ref{sec:theme3} traces a shift from implicit self-judgment during browsing to explicit and negotiable self-interpretation in conversation. Section~\ref{sec:theme4} examines the emotional consequences of these different sensemaking processes.
%% 下面呈现质性分析结果。同伴职业帖子使参与者能够接触到身边环境之外的职业路径，但他们仍需判断哪些案例与自己相关，以及这些经历对自己意味着什么。将帖子转化为可对话的 persona，改变了参与者浏览案例、形成问题以及把同伴经历关联到自身的方式。第~\ref{sec:theme1}~节和第~\ref{sec:theme2}~节说明参与者如何选择同伴案例并从中获得职业信息。第~\ref{sec:theme3}~节呈现从浏览时的内隐自我判断到对话中明确且可协商的自我解释。第~\ref{sec:theme4}~节考察这些不同意义建构过程所产生的情绪后果。

\subsection{Theme 1: Navigating Peer Career Posts}
\label{sec:theme1}

On RedNote, participants usually had to read a post before they could judge whether it was relevant. They opened a post, read its text or comments, and then returned to the post list. P13 described this as \emph{``reading one and then moving to the next.''} Although P13, P18, and P19 each viewed only around ten to fifteen posts, new posts continued to appear as they browsed, creating a sense of abundance. Because background, destination, and difficulties were not presented consistently, participants had to reconstruct a case before judging its relevance. They described the content as \emph{``messy,''} \emph{``too broad,''} \emph{``repetitive,''} or \emph{``reposted''} (P4, P10, P16, P18); P4 could not distinguish advertisements from genuine accounts. If the information they needed was missing, participants could only try to contact the author or search for another post. RedNote gave participants access to many career stories, but they had to identify relevant cases and decide where to look next on their own.
%% 在小红书上，参与者通常需要先阅读帖子，才能判断它是否与自己相关。他们打开帖子、阅读正文或评论，然后返回帖子列表。P13 将这种方式概括为\emph{``看完一个再看下一个''}。尽管 P13、P18 和 P19 各自只浏览了大约十到十五篇帖子，浏览过程中仍会不断出现新的帖子，使参与者感到内容十分丰富。由于背景、最终去向和困难没有被一致呈现，参与者必须先重构案例，才能判断其相关性。他们用\emph{``杂乱''}、\emph{``太宽泛''}、\emph{``重复''}或\emph{``转载''}描述这些内容（P4、P10、P16、P18）；P4 甚至无法区分广告和真实分享。如果所需信息有所缺失，参与者只能尝试联系作者或寻找另一篇帖子。小红书让参与者接触到许多职业故事，但他们必须自行识别相关案例，并决定下一步去哪里寻找信息。

By presenting posts as persona cards with consistent fields, JobMate allowed participants to judge whether a case was relevant before reading the full post. Participants could first exclude unsuitable cases and then open personas with a relevant background, destination, or difficulty. P21 said that visible backgrounds and destinations \emph{``help me filter faster and quickly pass over the ones that do not fit me''}; P2 found the tags intuitive, P9 said the cards reduced selection effort, and P24 could immediately skip cases of no interest. This compression was particularly valuable to P20, who found reading and remembering difficult, although the fast-reading P23 wanted more card detail. P7 considered the filtering effective, while P21 and P23 questioned the relevance or ordering of some recommendations. The structured fields helped participants compare cases and choose a persona. After making a choice, they could inspect the source post and ask about missing details. Transforming posts into personas thus shifted navigation from open-ended browsing toward case-based exploration. It did not remove concerns about card detail or recommendation clarity, but it made relevance cues visible earlier and reduced screening effort, consistent with the lower NASA-TLX effort reported for JobMate (Table~\ref{tab:quant-overview}).
%% JobMate 将帖子呈现为具有统一字段的 persona 卡片，使参与者可以在阅读完整帖子之前判断一个案例是否相关。参与者可以先排除不合适的案例，再打开在背景、去向或困难方面与自己相关的 persona。P21 表示，可见的背景和去向\emph{``能帮助自己更快地筛选，把和自己不合适的很快就 pass 掉''}；P2 认为标签直观，P9 表示卡片降低了选择成本，P24 则可以立即跳过不感兴趣的案例。对于阅读和记忆较困难的 P20，这种压缩尤其有帮助；但阅读较快的 P23 希望卡片提供更多细节。P7 认为筛选机制有效，而 P21 和 P23 对部分推荐的相关性或排序有所疑问。结构化字段帮助参与者比较案例并选择一个 persona。作出选择后，他们可以查看原帖，并继续询问缺失的信息。因此，将帖子转化为 personas，使 navigation 从开放式浏览转向基于案例的探索。它并未消除卡片细节或推荐清晰度方面的问题，但使相关性线索更早可见并降低了筛选成本，这与 JobMate 组较低的 NASA-TLX 努力程度评分一致（表~\ref{tab:quant-overview}）。

\subsection{Theme 2: Information Gained from Peer Posts}
\label{sec:theme2}

Peer posts made career possibilities more concrete, but the paths participants saw depended on which experiences were available. Participants learned which paths people with related backgrounds had taken. On RedNote, participants often had to find a person's background and career destination across the post and its comments. JobMate presented this information in fixed persona fields. In both conditions, P6 learned that psychology graduates could enter user research, P12 found the \emph{``real career destinations''} of people from the same major, P15 reconsidered a path previously dismissed, and P24 encountered options beyond everyday experience. However, the available posts did not cover all career paths. P22 saw mostly big-tech trajectories despite the prevalence of public-sector careers around them, and P23 identified the same gap in JobMate; P1 and P23 also wanted more detail. JobMate made the paths in its collection easier to recognise, but it could not show paths that were absent from that collection.
%% 同伴帖子使职业可能性变得更加具体，但参与者能看到哪些路径，取决于现有帖子提供了哪些经历。参与者从中了解到，具有相似背景的人选择了哪些职业路径。在小红书上，参与者往往需要从帖子正文和评论中寻找作者的背景与职业去向。JobMate 则使用固定的 persona 字段呈现这些信息。在两种条件下，P6 了解到心理学毕业生可以从事用户研究，P12 看到了同专业者的\emph{``真实的职业去向''}，P15 重新考虑了此前排除的路径，P24 则接触到日常生活之外的选择。然而，现有帖子并未涵盖所有职业路径。尽管 P22 身边许多人选择体制内职业，浏览到的却主要是大厂路径；P23 在 JobMate 中也发现了相同的缺口；P1 和 P23 还希望看到更多细节。JobMate 使其语料中的职业路径更容易被识别，但无法呈现语料中缺少的路径。

Participants also used peer posts to learn what particular jobs were like and how to prepare for them. The two conditions differed in how participants found and used these details. On RedNote, participants found information that changed their expectations. P19 found the threshold for algorithm jobs \emph{``not as high as I had imagined,''} and P17 identified application steps and gaps in their resume. However, participants had to locate such details within posts that were repetitive, mixed with advertisements, or too technical to understand (P10, P16). With JobMate, participants could ask follow-up questions about missing details. Initial answers were sometimes general when users provided little personal information (P3, P7, P23, P24), but became more relevant when users described their own constraints. P11's broad question about finding a first internship initially produced several types of organisations. After adding \emph{``Guangzhou, weekdays except Wednesday,''} the options narrowed to locally feasible settings and eventually led to a request for a \emph{``Guangzhou psychology internship information checklist.''} Answers in other conversations also became more specific when participants described their abilities or eligibility requirements. RedNote offered detailed but noisy posts, while JobMate let participants ask directly about missing information and relate general information to their own situations. Both conditions reduced participants' reported lack of career information and received similar information-support ratings (Table~\ref{tab:quant-overview}). The main difference was how participants accessed and used the information, rather than how much information they felt they had received.
%% 参与者还通过同伴帖子了解具体工作是什么样的，以及应该如何为这些工作做准备。两种条件的区别在于参与者如何找到和使用这些具体信息。在小红书上，参与者找到了能够改变原有预期的信息。P19 发现算法岗位的门槛\emph{``没有自己想象的那么高''}，P17 则明确了求职步骤和简历中的不足。然而，参与者必须从重复、夹杂广告或技术内容过于晦涩的帖子中寻找这些细节（P10、P16）。在 JobMate 中，参与者可以针对缺失的细节继续提问。当用户提供的个人信息较少时，最初的回答有时较为宽泛（P3、P7、P23、P24）；但当用户说明自己的限制条件后，回答会变得更贴合其情况。P11 最初笼统地询问如何寻找第一段实习，得到的是几类机构。在补充\emph{``广州，工作日除周三''}后，选项收窄到本地可行的场所，并最终形成了对一份\emph{``广州心理实习找信息清单''}的需求。在其他对话中，当参与者说明自己的能力或资格要求后，回答也变得更加具体。小红书提供了详细但嘈杂的帖子，而 JobMate 让参与者能够直接询问缺失的信息，并把一般信息联系到自己的情况。两种条件都降低了参与者报告的信息缺乏程度，并获得了相近的信息支持评分（表~\ref{tab:quant-overview}）。主要区别在于参与者如何获取和使用信息，而不是他们认为自己获得了多少信息。

\subsection{Theme 3: From Implicit Comparison to Negotiable Self-Interpretation}
\label{sec:theme3}

Transforming posts into conversational personas changed how participants connected peer experiences to their own situations. On RedNote, participants formed implicit self-judgments against standards accumulated across posts, but these judgments remained private and unavailable for response. In JobMate, dialogue first prompted users to describe circumstances and criteria that had remained implicit. The agent then organised this information and made statements about the users' abilities, experiences, or career fit. Users could agree with, question, or correct these statements. Participants did not always respond to or revise the agent's statements. We observed these responses mainly when participants provided concrete personal details, the agent addressed those details directly, and participants continued the discussion.
%% 将帖子转化为对话式 persona，改变了参与者把同伴经历与自身处境联系起来的方式。在小红书上，参与者依据许多帖子累积形成的标准作出内隐自我判断，但这些判断保持私人化，也无法获得回应。在 JobMate 中，对话首先促使用户描述此前保持内隐的处境与标准；agent 随后组织这些信息，并对用户的能力、经历或职业适配性作出描述。用户可以认同、质疑或纠正这些描述。参与者并不总会回应或修订 agent 的说法。我们主要在参与者提供具体个人信息、agent 直接回应这些信息，并且参与者继续讨论时观察到这些回应。

\subsubsection{Implicit Self-Judgment While Browsing RedNote}

RedNote prompted implicit self-judgment by repeatedly showing what other jobseekers had achieved and how they had prepared. Because these reference points appeared across many posts, participants measured themselves against an aggregated image of ``others'' rather than one clearly defined peer. No specific speaker could be asked to explain or revise this standard. Seven RedNote participants used ``others'' in twelve statements about comparison. P16 described seeing that \emph{``others possess abilities needed for a position, while I do not,''} while P12 focused on \emph{``others doing so many internships in the same few years and reaching a good destination.''} These judgments sometimes created direction, but the resulting clarity was unstable. P10 explained, \emph{``It feels clear when I finish reading, but after a while it becomes unclear again---the same as before I browsed.''} P13 similarly concluded that browsing RedNote \emph{``cannot help with the choice,''} while P22 had to process conflicting views from people with different educational backgrounds \emph{``in my own head.''} Browsing prompted reflection but did not require participants to state what they had accepted, rejected, or learned about themselves. They therefore had no explicit judgment to revisit or speaker to answer. Their self-judgment remained implicit, while the clarity it produced often faded after browsing.
%% 小红书通过反复展示其他求职者取得了什么结果以及如何准备，触发了内隐的自我判断。由于这些参照分散在许多帖子中，参与者衡量自己时所面对的是一个由许多内容聚合而成的“别人”，而不是某个明确的同伴，也没有具体的说话者可以被要求解释或修改这一标准。小红书组有七位参与者在十二句比较相关的表述中使用了“别人”。P16 描述自己看到\emph{``别人具备进入某个岗位所需的能力，而自己不具备这些能力''}；P12 则关注到\emph{``别人在相同的几年时间实习了很多次，并获得了很好的去向''}。这些判断有时会带来方向感，但由此产生的清晰感并不稳定。P10 表示：\emph{``看完的时候觉得清晰，过一会儿又不清晰了，和没刷的时候感觉一样。''}P13 同样认为信息流\emph{``不能在选择上给出帮助''}；P22 则只能\emph{``自己用脑子''}处理来自不同教育背景用户的冲突观点。浏览会触发反思，却不会要求参与者说出自己接受、拒绝了什么，或者形成了怎样的自我认识。因此，他们既没有明确判断可以回看，也没有说话者可以回应。自我判断保持内隐，而它产生的清晰感往往会在浏览结束后消散。

\subsubsection{Talking Through Preferences Made Career Criteria Clearer}

Dialogue turned unclear preferences into explicit criteria by requiring users to describe relevant experiences, abilities, concerns, and unacceptable working conditions. The agent accumulated and restated these details for users to confirm or refine. P5 described this mechanism directly: \emph{``Chatting requires output, which helps me absorb information better and completes the loop of understanding.''} P2's conversation shows how saying what one did not want could define what might fit. They initially rejected none of the suggested directions, but later stated, \emph{``I do not really want to do counselling; the emotional burden is too heavy,''} and \emph{``I do not like work with high analysis or data demands.''} The agent gathered these exclusions into three conditions: no counselling, willingness to work with people, and no sustained high emotional load. P2 then asked specifically about employee care. Other participants clarified choices by identifying their underlying trade-offs. P9 moved from comparing functional roles in state-owned enterprises with market or business roles to explaining a fear of being pushed, bearing responsibility for mistakes, and facing high uncertainty. P23 reframed teaching in Hong Kong or Shenzhen as a trade-off involving living costs and housing, and remained undecided. Rather than necessarily producing a final choice, these exchanges turned ``what suits me'' into conditions participants could examine.
%% 对话要求用户描述相关的经历、能力、担忧以及无法接受的工作条件，从而把模糊偏好转化为明确标准。Agent 会积累并复述这些细节，供用户确认或调整。P5 直接描述了这一机制：\emph{``聊天需要输出，能够更好地吸收信息，完成信息理解的闭环。''}P2 的对话展示了，说出自己不想要什么可以怎样界定可能适合自己的选择。最初，他们不排斥建议的任何方向，随后却逐渐说出：\emph{``不是很想做咨询，情绪负担太重了''}，以及\emph{``不太喜欢分析或者数据要求太高的工作''}。Agent 将这些排除项整理为三个条件：不做咨询、愿意与人打交道、不长期承受高情绪负荷。之后，P2 具体询问了员工关怀。其他参与者则通过指出选择背后的取舍来澄清问题。P9 从比较央国企职能岗与市场或商务岗，逐渐转向说明自己担心被人 push、为失误承担责任以及面对高度不确定。P23 则把去香港还是深圳当老师重新表述为生活成本和住房方面的权衡，并且最终仍未作出决定。这些交流未必产生最终选择，却把“什么适合我”转化成了参与者可以检查的具体条件。

\subsubsection{Accepting, Challenging, and Revising Agent Statements About Skills and Career Fit}

JobMate allowed participants to accept, challenge, or revise what the agent said about their skills, experiences, and career fit. The agent described what participants' experiences might show about their skills, assessed the value of those experiences, or suggested jobs that might suit them. Participants responded to particular descriptions of themselves rather than simply accepting or rejecting an entire response. P21 accepted a new view of a project, reporting that the conversation \emph{``changed how I viewed a project that I had thought was a waste of time.''} P14 likewise used the agent to describe previous activities as abilities that could be presented for a target role. P1, however, challenged an overly positive description of their experience. The agent described P1's paper as \emph{``more solid than a pile of so-called projects''} and claimed that it showed an understanding of LLMs and RAG. P1 corrected the account: \emph{``Actually, I only called a framework for RAG; I do not really understand RAG techniques, and I have never worked on back-end development.''} P1 was not simply persuaded by the agent. By correcting this overstatement, they produced a more precise description of their technical boundaries and asked what skills the target role required. P5 demonstrated revision and renewed acceptance. They initially devalued an internship because \emph{``most of the time was spent cleaning data,''} but later asked the agent to rewrite it for a resume and interview while narrowing the target to foundation-model post-training at a big tech company. The usable account was neither the user's initial dismissal nor the agent's unrestricted praise, but a version revised through interaction. JobMate therefore supported reflection by allowing users to agree with, challenge, and revise what the agent said about them, rather than presenting an assessment they had to accept.
%% JobMate 使参与者能够接受、质疑或修订 agent 对其能力、经历和职业适配性的说法。Agent 会说明参与者的经历可能体现哪些能力、判断这些经历的价值，或建议可能适合他们的工作。参与者会回应关于自己的具体描述，而不是简单地接受或拒绝整段回答。P21 接受了对一个项目的新看法，并表示对话\emph{``扭转了对于自己觉得浪费时间的项目的看法''}。P14 同样利用 agent，把过去的活动描述为可以面向目标岗位呈现的能力。P1 则质疑了对其经历过于积极的描述。Agent 将 P1 的论文描述为\emph{``比一堆所谓的项目硬多了''}，并认为这说明 P1 懂 LLM 和 RAG。P1 纠正道：\emph{``其实我对 RAG 只是调用了一下框架，我不是很了解 RAG 的技巧那些……还有后端开发我也完全没有接触过。''}P1 并不是简单地被 agent 说服。通过纠正这一夸大的说法，他们形成了对自身技术边界更准确的描述，并进一步询问目标岗位需要哪些技能。P5 则展示了修订后重新接受的过程。他们最初因为\emph{``大部分时间都是在清洗数据''}而否定一段实习的价值，之后却要求 agent 将它改写为简历和面试表述，同时把目标收窄到大厂的大模型后训练实习。最终可用的表述既不是用户最初的完全否定，也不是 agent 不加限制的肯定，而是双方在互动中修订出的版本。因此，JobMate 支持反思的方式不是给出用户必须接受的评价，而是让用户能够认同、质疑和修订 agent 对自己的说法。

\subsubsection{Users Responded When the Agent Used User-Provided Details}

Participants were better able to respond when the agent clearly addressed their abilities, experiences, or career fit and they continued the conversation. On RedNote, the standards formed across many posts had no clear speaker whom users could answer. A general AI response came from a clear speaker, but users still had little to accept, challenge, or revise if the response did not address their actual abilities, experiences, or constraints. In P1's case, the agent made a clear statement about P1's RAG ability, and P1 explained exactly why it was inaccurate. When replies remained broad, the interaction developed differently. P7 considered the summaries too general to support further conversation, while P23 and P24 also found some responses broad. Continued conversation mattered as well. P3 ended an exchange soon after saying, \emph{``I am afraid I am no good at any of them,''} without developing the concern further. P20 mainly asked about resumes and interviews, and P24 focused on costs, privacy, and persona backgrounds. These practical uses still had value, but they did not require participants to discuss what the agent said about them. JobMate therefore gave participants a chance to express and revise their judgments, but it did not automatically lead to reflection. Participants could examine and revise the agent's statements when those statements addressed specific abilities, experiences, or constraints and when participants had provided enough personal detail to respond.
%% 当 agent 明确回应参与者的能力、经历或职业适配性，并且参与者继续对话时，他们更容易作出回应。在小红书上，由许多帖子共同形成的标准缺少一个可以被用户回应的明确说话者。一般性的 AI 回答虽然来自一个明确的说话者，但如果没有涉及用户实际的能力、经历或限制，用户仍然缺少可以接受、质疑或修订的内容。在 P1 的案例中，agent 明确评价了 P1 的 RAG 能力，而 P1 说明了这一说法具体错在哪里。当回复保持宽泛时，互动会以不同方式发展。P7 认为总结过于一般，无法支持继续交流；P23 和 P24 也认为部分回答较为宽泛。继续对话同样重要。P3 在说出\emph{``怕自己哪都不行''}后不久便结束了一次交流，没有进一步展开这一担忧。P20 主要询问简历和面试，P24 则集中了解费用、隐私和 persona 背景。这些实际用途仍然具有价值，但不需要参与者讨论 agent 对自己的说法。因此，JobMate 为参与者提供了表达和修订自身判断的机会，但不会自动带来反思。当 agent 的说法涉及具体的能力、经历或限制，并且参与者已经提供了足够的个人信息来回应时，他们可以检查并修订这些说法。

Overall, RedNote and JobMate differed in how participants could respond to social comparisons and to what the agent said about them. RedNote prompted reflection through largely implicit comparison with an accumulated image of ``others.'' Persona-agent interaction instead linked three observable actions: users described their personal conditions, the agent organised this information and responded to it, and users accepted, challenged, or revised the agent's statements. Both conditions showed lower mean career decision-making difficulty after exploration and similar mean self-determination scores (Table~\ref{tab:quant-overview}). The contribution of conversation was not a larger descriptive decrease in difficulty, but a form of sensemaking in which users' reasoning became visible, responsive, and revisable. This form was observed mainly when users gave enough personal context and the agent responded directly to it.
%% 总体而言，小红书和 JobMate 的区别在于参与者如何回应社会比较，以及如何回应 agent 对自己的说法。小红书主要通过用户与许多帖子累积形成的“别人”进行内隐比较来触发反思。Persona-agent 互动则连接了三个可观察的动作：用户描述个人条件，agent 组织并回应这些信息，而用户接受、质疑或修订 agent 的说法。两种条件在探索后的平均职业决策困难均有所降低，整体自我决定平均评分也较为接近（表~\ref{tab:quant-overview}）。对话的贡献并不是描述性数据中更大的困难下降，而是形成一种使用户推理变得可见、可以被回应并能够被修订的意义建构方式。我们主要在用户提供足够的个人情境，并且 agent 直接回应这些信息时观察到了这种形式。

\subsection{Theme 4: Emotional Consequences of Different Sensemaking Processes}
\label{sec:theme4}

RedNote browsing could both reassure participants and make them more anxious. Peer cases served simultaneously as reassurance and as standards for self-evaluation. Similar uncertainty normalised difficulty: P13 felt less anxious after seeing people from the same major express anxiety, while P18 said that learning about others' psychological changes during applications \emph{``reduced the pressure.''} Yet more successful peers exposed perceived gaps. P16 worried about abilities they lacked, P12 became more anxious after seeing others complete many internships and reach good destinations, and P17 focused on how much remained to learn. P6 captured both consequences: \emph{``Knowing that others are as lost as I am makes me less worried, but seeing people more competitive than me makes me anxious.''} Reassurance and anxiety thus came from the same comparative process, and the reference points changed from one post to the next, making reassurance unstable. P16 consequently did not feel understood or accompanied, P4 relied more on family and friends, and P22 found the experiences insufficiently real.
%% 浏览小红书既可能让参与者感到安慰，也可能使他们更加焦虑。同伴案例既提供安慰，也成为自我评价的标准。相似的不确定感使困难显得正常：P13 看到同专业的人表达焦虑后减轻了焦虑，P18 则表示，了解他人求职过程中的心理变化\emph{``减轻了压力''}。然而，更成功的同伴也会暴露参与者感知到的差距。P16 担忧自己缺少某些能力，P12 在看到他人完成多段实习并获得较好去向后更加焦虑，P17 则关注到自己还有很多内容需要学习。P6 同时概括了两种后果：\emph{``知道别人和自己一样迷茫就会不那么担心，但是看到竞争力比自己强的人会焦虑。''}因此，安慰和焦虑来自同一个比较过程，而参照对象随着帖子不断变化，使安慰难以稳定。P16 因而没有感到被理解或陪伴，P4 更多依赖家人朋友，P22 则认为这些经历不够真实。

JobMate made reassurance more direct by addressing the user and showing difficulties alongside outcomes on persona cards. P9 described a \emph{``sense of participation''} and immediate affirmation; P14 reported feeling understood and accompanied; and even the critical P23 valued being told that the anxiety was not solely their problem and \emph{``felt that I was not alone.''} P20 likewise found that challenge tags and seeing how long others had applied reduced anxiety. However, the same enabling conditions identified in Section~\ref{sec:theme3} also governed emotional support: it was credible when tied to recognisable difficulties and concrete evidence, not merely when positive. P24 described a general response as \emph{``a little like chicken soup for the soul,''} P23 found some replies vague, and P1 corrected an inflated interpretation of their abilities.
%% JobMate 通过直接回应用户，并在 persona 卡片上同时展示困难和结果，使安慰变得更加直接。P9 描述了\emph{``参与感''}和即时肯定；P14 表示感到被理解和陪伴；即使是较为挑剔的 P23，也认可 agent 告诉自己焦虑不全是个人的问题，并因此\emph{``觉得自己不是一个人''}。P20 同样认为，困难标签以及看到他人求职所花费的时间有助于减轻焦虑。不过，第~\ref{sec:theme3}~节中发现的同样促进条件也决定了情绪支持是否有效：支持只有在联系到可辨认的困难和具体证据时才显得可信，而不是只要积极就足够。P24 将一般性的回答形容为\emph{``有点心灵鸡汤''}，P23 认为部分回复空泛，P1 则纠正了 agent 对自身能力的夸大解释。

Mean perceived emotional-support ratings were close across conditions (Table~\ref{tab:quant-overview}). The qualitative difference therefore concerned how emotion arose within sensemaking: RedNote tied it to a changing sequence of comparisons, whereas JobMate made support a direct but contestable response. Across both conditions, emotional support depended on whether users recognised it as specific, credible, and relevant to their present situation.
\section{Discussion}
\label{sec:discussion}

Our comparison reveals how two interaction forms organise the interpretive work of using peer experience differently. RedNote exposed participants to numerous concrete trajectories but left them to reconstruct cases and privately judge their relevance. JobMate bounded exploration around identifiable peer cases and made some of that judgment explicit in dialogue. We discuss this transformation in relation to possible selves, social comparison and identity work, and prior systems for making user-generated experiences usable.
%% 我们的比较揭示了两种交互形式如何以不同方式组织使用同伴经历时所需的解释工作。小红书让参与者接触到大量具体轨迹，却由他们自行重构案例并私下判断其相关性；JobMate 则围绕可识别的同伴案例限定探索范围，并通过对话使部分判断变得明确。下面我们联系 possible selves、社会比较与身份工作，以及既有的用户生成经验支持系统，讨论这一转变。

\subsection{From Possible Career Selves to Negotiable Self-Interpretations}
\label{subsec:disc-selves}

Peer career posts are valuable because they give possible futures a concrete social form. Career-exploration research distinguishes exploration of the self from exploration of educational and occupational environments~\cite{Stumpf1983CareerExploration,Flum2000VocationalExploration,Jiang2019CareerExploration}; peer posts bring both into the same encounter. Prior work shows that young job seekers use social media to encounter trajectories outside their immediate surroundings and imagine possible selves~\cite{Corvite2024,Mowbray2021,Markus1986}. Ibarra's account of provisional selves further suggests that professional identity develops through trying out tentative versions of who one might become~\cite{Ibarra1999}. Our findings add a process account to this literature. Participants did not treat a peer's destination as a possibility in the abstract; they reconstructed the person's background, difficulties, and outcome, then asked whether the path could apply to someone in their own circumstances. The same specificity that made a future imaginable also made it necessary to interpret the distance between the peer and the self.
%% 同伴职业帖子的价值在于，它们为可能的未来赋予了具体的社会形式。既有研究表明，年轻求职者通过社交媒体接触身边环境之外的职业轨迹，并想象 possible selves~\cite{Corvite2024,Mowbray2021,Markus1986}。Ibarra 关于 provisional selves 的论述进一步指出，职业身份会在尝试“自己可能成为谁”的暂时版本中发展~\cite{Ibarra1999}。我们的发现为这条文献补充了过程层面的解释。参与者并不会把某位同伴的去向仅仅视为抽象可能性；他们会重构对方的背景、困难和结果，再判断这条路径是否适用于自身处境。使未来变得可想象的具体性，也使用户必须解释同伴与自我之间的距离。

This interpretive step extends prior HCI work on sensemaking with community content. Systems such as ComViewer, PlanHelper, Sensecape, and Graphologue help users filter, extract, restructure, or comprehend fragmented information~\cite{Wu2025ComViewer,Li2022PlanHelper,Suh2023Sensecape,Jiang2023Graphologue}; career systems similarly organize job, skill, and trajectory information for exploration~\cite{Ashok2018JobSense,JoseGarcia2023C3IoC}. Conversational search further lets users query community content rather than only browse it~\cite{Wu2026ConSearcher}. These systems primarily make content more intelligible. JobMate indicates that career exploration contains an additional transition: after understanding a peer case, the user must decide what to take from it as a claim about their own possibilities. Conversation supported this transition not merely because it supplied answers, but because asking and responding required users to state constraints, exclusions, and trade-offs. This is consistent with work showing that constructive and interactive engagement can deepen processing~\cite{Chi2014}, while specifying that, here, the relevant output was an articulated relation between a peer trajectory and the self.
%% 这一解释步骤扩展了 HCI 中关于利用社区内容进行意义建构的研究。ComViewer、PlanHelper、Sensecape 和 Graphologue 等系统帮助用户筛选、提取、重组或理解碎片化信息~\cite{Wu2025ComViewer,Li2022PlanHelper,Suh2023Sensecape,Jiang2023Graphologue}；对话式搜索则使用户能够询问社区内容，而不只是浏览~\cite{Wu2026ConSearcher}。这些系统主要使内容变得更可理解。JobMate 表明，职业探索还包含一个额外的转变：理解同伴案例之后，用户还必须判断自己应当从中接受哪些关于自身可能性的说法。对话之所以支持这一转变，不只是因为它提供回答，还因为提问和回应要求用户说出限制、排除项和取舍。这与交互式参与能够深化信息加工的研究一致~\cite{Chi2014}，同时进一步说明：在本研究中，关键产出是同伴轨迹与自我之间被明确表达的关系。

This direction of interpretation distinguishes JobMate from two prominent forms of career agent. Normative counselling agents guide choices, while future-self agents project the user forward to support reflection and continuity~\cite{Gati2021CareerDecisions,Lee2021CASExplorer,Han2025SDT,Du2024CareerSim,Jeon2025LettersFutureSelf,Pataranutaporn2024FutureYou}. Persona systems can add an external standpoint, but often through fictional experts, generated positions, or simulated characters~\cite{Liu2025PersonaFlow,Park2023GenerativeAgents}. A post-grounded persona occupies a different position: it is neither an authoritative counsellor nor a projection of the user, but a mediated standpoint anchored in what a particular peer publicly experienced. When that standpoint interpreted the user's abilities or fit, the peer experience appeared to ``speak back.'' Our data do not show that these interpretations were more accurate or led to better decisions. They show that the transformation created an encounter in which users could articulate why a peer case did or did not apply to them and respond to an interpretation rather than leave it private.
%% 这种解释方向使 JobMate 区别于两类典型的职业 agent。规范性咨询 agent 引导用户作出选择，而未来自我 agent 则把用户投射到未来，以支持反思和自我连续性~\cite{Han2025SDT,Du2024CareerSim,Jeon2025LettersFutureSelf,Pataranutaporn2024FutureYou}。Persona 系统可以引入外部立场，但通常通过虚构专家、生成的观点或模拟角色来实现~\cite{Liu2025PersonaFlow,Park2023GenerativeAgents}。基于原帖的 persona 占据了不同的位置：它既不是权威咨询者，也不是用户自身的投射，而是锚定在某位具体同伴公开经历上的中介立场。当这一立场解释用户的能力或适配性时，同伴经历仿佛“反过来对用户说话”。我们的数据不能说明这些解释更加准确或带来了更好的决策；它表明，这种转化创造了一种相遇，使用户能够说出某个同伴案例为什么适用或不适用于自己，并回应一种解释，而不是让判断一直保持私人化。

\subsection{Making Peer Comparison Explicit Without Making It Normative}
\label{subsec:disc-comparison}

Social comparison research explains why peer trajectories can support and unsettle users at the same time: comparison supplies information for self-evaluation, while upward comparison can motivate or diminish the self depending on how it is interpreted~\cite{Festinger1954,Buunk1990,Appel2023}. Career-related studies have documented its association with pressure, frustration, and anxiety~\cite{Fukubayashi2021,Zhang2024Pressure,Luo2024}. Our findings locate an interactional difference within this process. During browsing, participants compared themselves with an accumulated and shifting image of ``others.'' The standard was consequential but had no single speaker whom a user could question. In JobMate, an interpretation such as whether an experience demonstrated technical ability or fit a role was uttered by a definite conversational counterpart. This did not remove comparison; it gave the comparison an addressable form.
%% 社会比较研究解释了为什么同伴轨迹能够同时支持用户并使其不安：比较为自我评价提供信息，而向上比较会根据其解释方式激励或削弱自我~\cite{Festinger1954,Buunk1990,Appel2023}。职业相关研究已经记录了社会比较与压力、挫败感和焦虑之间的关联~\cite{Fukubayashi2021,Zhang2024Pressure,Luo2024}。我们的发现进一步定位了这一过程中的交互差异。在浏览中，参与者把自己与一个不断累积、持续变化的“别人”进行比较。这一标准会产生影响，却没有一个可以被用户追问的单一说话者。在 JobMate 中，某段经历是否体现 技术能力、某人是否适合某个岗位等解释，则由一个明确的对话对象说出。对话并未消除比较，而是赋予了比较一种可以被回应的形式。

Identity work offers a vocabulary for understanding what participants then did. Identities can be \emph{ascribed} by others and \emph{avowed} through one's own talk~\cite{Collier1988,Snow1987}; early professional identity likewise develops by testing provisional selves against external feedback~\cite{Ibarra1999}. We use these concepts descriptively rather than treating the agent as an identity authority. JobMate produced ascriptions about ability, experience value, or occupational fit, but participants did not simply accept them. P21 accepted a reinterpretation of a previously devalued project; P1 challenged an inflated account of their RAG expertise and articulated more precise technical boundaries; P5 revised both their own dismissal of an internship and the agent's positive framing before reusing the experience. Acceptance, challenge, and revision were therefore forms of identity work through which users negotiated what could count as a valid description of themselves.
%% 身份工作为理解参与者随后所做的事情提供了一组词汇。身份既可能被他人\emph{指认}，也可能通过个人自己的言说被\emph{认领}~\cite{Collier1988,Snow1987}；职业早期身份同样会在用外部反馈检验 provisional selves 的过程中发展~\cite{Ibarra1999}。我们以描述性方式使用这些概念，而不把 agent 视为身份权威。JobMate 会提出有关能力、经历价值或职业适配性的指认，但参与者并非简单接受。P21 接受了对一段此前被低估项目的新解释；P1 反驳了对其 RAG 专长的夸大描述，并说出了更准确的技术边界；P5 则同时修订了自己对实习的否定和 agent 的积极框定，之后才重新使用这段经历。因此，接受、质疑与修订构成了身份工作的不同形式，用户借此协商什么可以成为对自己的有效描述。

Making comparison explicit should not make the agent's interpretation normative. SDT-based career systems show how autonomy, competence, and relatedness can support decision processes without prescribing a single correct choice~\cite{Ryan2000,Han2025SDT}. In our data, autonomy was visible not as compliance but as participants' ability to decline a suggestion, state an exclusion, or correct the agent. This matters because conversational fluency can otherwise make a generated interpretation sound more authoritative than the peer evidence warrants. A useful peer-experience agent should therefore present interpretations as provisional propositions---``does this fit your situation?''---rather than verdicts about who the user is or what they should do. Challenge is not a breakdown of the interaction; it is evidence that the user retains authorship over the self-description being constructed.
%% 使比较显性化，并不意味着让 agent 的解释具有规范性。基于 SDT 的职业系统说明，autonomy、competence 和 relatedness 可以在不规定唯一正确选择的情况下支持决策过程~\cite{Ryan2000,Han2025SDT}。在我们的数据中，autonomy 并不表现为服从，而表现为参与者能够拒绝建议、说出排除项或纠正 agent。这一点很重要，因为流畅的对话可能让生成式解释听起来比其同伴证据所能支持的更加权威。因此，有用的同伴经验 agent 应把解释呈现为暂时的命题——“这是否符合你的处境？”——而不是关于用户是谁或应该做什么的裁决。质疑不是互动的失败，而是用户仍然保有对正在构建的自我描述之作者权的证据。

This opportunity was also uneven. Articulating one's situation is itself substantive work, related to the effects of self-disclosure and being responded to~\cite{Ho2018,Collins1994}. Participants who could name locations, time constraints, disliked work conditions, or technical limits gave the agent material from which it could form a specific and contestable interpretation. Participants who supplied little context, or received generic replies, had less to accept or challenge and often disengaged. This echoes difficulties reported for conversational community search, where persona selection and over-personalisation can impede engagement~\cite{Wu2026ConSearcher}. The implication is not that users who articulate less cannot reflect, but that conversation redistributes work: it makes interpretation observable while depending on users' capacity and willingness to put their circumstances into words. Personalisation should therefore be treated as a supported interactional achievement, not an automatic property of dialogue.
%% 这种机会也并非均等出现。表达自身处境本身就是一项实质性工作，并与自我披露和获得回应的作用有关~\cite{Ho2018,Collins1994}。能够说出地点、时间限制、不喜欢的工作条件或技术边界的参与者，为 agent 提供了形成具体且可质疑解释的材料。提供情境较少或只得到通用回复的参与者，则缺少可以接受或质疑的内容，也更容易停止互动。这与社区对话式搜索中报告的困难相呼应：persona 选择和过度个性化都可能妨碍参与~\cite{Wu2026ConSearcher}。这并不意味着表达较少的用户无法反思，而是说明对话重新分配了解释工作：它使解释变得可观察，同时依赖用户把自身处境说出来的能力和意愿。因此，个性化应当被视为一种需要支持的互动成就，而不是对话自动具备的属性。

\subsection{Designing Source-Traceable and Contestable Peer-Experience Agents}
\label{subsec:disc-design}

Our findings suggest that systems mediating peer experience should preserve a visible distinction among three layers: what the original peer wrote, what additional community materials contribute, and what the model infers about the current user. Prior systems improve access to community knowledge through filtering and synthesis~\cite{Wu2025ComViewer,Li2022PlanHelper}, but generative synthesis can narrow or smooth the diversity of source material~\cite{Sharma2024EchoChamber}. In JobMate, keeping the original post beside the conversation gave users a potential route back to the lived account. Future designs should make this route more actionable by attaching generated claims to source passages, indicating when a response draws on retrieved posts rather than the core persona, and visually separating source-supported statements from model interpretation. Source visibility should support inspection, not merely serve as a credibility cue.
%% 我们的发现表明，中介同伴经历的系统应当在三个层次之间保持可见区分：原始同伴写了什么、补充的社区材料提供了什么，以及模型对当前用户推断了什么。既有系统通过筛选和综合改善社区知识的可访问性~\cite{Wu2025ComViewer,Li2022PlanHelper}，但生成式综合可能缩窄或平滑来源材料的多样性~\cite{Sharma2024EchoChamber}。在 JobMate 中，将原帖保持在对话旁边，为用户提供了一条返回真实经历的潜在路径。未来设计应通过把生成的说法关联到具体原文、标明回答何时来自检索帖子而非核心 persona，并在视觉上区分来源支持的陈述与模型解释，使这条路径更可操作。来源可见性应当支持检查，而不应只是增加可信感的提示。

Second, systems should scaffold articulation rather than require users to arrive with a complete self-description. Our participants often found it easier to state what they did not want than to specify an ideal career. Interfaces could therefore elicit exclusions before preferences, let users edit a compact profile of constraints across turns, and show how each disclosed detail changes a recommendation or interpretation. Matching should likewise be explainable: users should see whether a persona was surfaced because of background, destination, or shared difficulty and be able to change that basis. These mechanisms extend the information-organisation goals of prior sensemaking tools~\cite{Suh2023Sensecape,Jiang2023Graphologue} from structuring external content to structuring the user's evolving account of their own situation.
%% 第二，系统应为表达提供支架，而不是要求用户一开始就带着完整的自我描述。我们的参与者常常发现，说出自己不想要什么，比说明理想职业更容易。因此，界面可以先引出排除项，再询问偏好；允许用户跨轮次编辑一份简洁的个人限制档案；并展示每一项披露的信息如何改变推荐或解释。匹配过程也应当可解释：用户需要知道某个 persona 是因为背景、去向还是共同困难而被呈现，并且能够改变这一匹配依据。这些机制把既有意义建构工具组织外部信息的目标~\cite{Suh2023Sensecape,Jiang2023Graphologue}，扩展到了组织用户不断发展的自我处境叙述。

Third, generated interpretations should be specific enough to inspect and deliberately easy to contest. Specificity gives the user something concrete to evaluate, but it also increases the risk of overclaiming. Designs could phrase interpretations with calibrated uncertainty, expose the evidence used, invite correction, and preserve the user's correction as part of subsequent context. The goal is not to maximise agreement or positive affect. As our participants' responses to vague encouragement and inflated praise show, support was credible when it referred to recognisable evidence and remained open to revision. Classical distinctions between source and message credibility are useful here~\cite{Hovland1951}: grounding identifies where an experience came from, but does not by itself validate the model's new claim about the user.
%% 第三，生成的解释应当足够具体以供检查，同时被有意设计为容易质疑。具体性让用户拥有可评价的明确对象，却也增加了过度断言的风险。系统可以使用经过校准的不确定性表达、展示所依据的证据、主动邀请纠正，并把用户的纠正保留为后续情境的一部分。目标不是最大化同意或积极情绪。正如参与者对宽泛鼓励和夸大奖励的反应所示，只有当支持指向可识别的证据并保持可修订时，它才显得可信。关于来源可信度与信息可信度的经典区分在这里仍然有用~\cite{Hovland1951}：grounding 能说明一段经历来自哪里，却不能自动验证模型针对用户提出的新说法。

Persona distinctiveness is a secondary but consequential design choice within this framework. Prior persona systems often cultivate vivid characters or stances~\cite{Ha2024CloChat,Liu2025PersonaFlow,Park2023GenerativeAgents}, whereas JobMate grounded personas in experience without predefined personality. This kept attention on trajectories and user circumstances, but participants sometimes perceived different personas as speaking with the same voice. Simply adding invented personality would weaken source fidelity and could encourage users to attribute generated behaviour to the original author. A more cautious middle path is to preserve limited stylistic cues already present in the post---for example, whether its tone is instructional, matter-of-fact, or self-deprecating---while labeling them as source-derived and avoiding unsupported character traits. This treats formal fidelity as part of grounding without equating the conversational simulation with the person who wrote the post.
%% Persona 的辨识度是这一框架下次要但仍有影响的设计选择。既有 persona 系统通常塑造鲜明的角色或立场~\cite{Ha2024CloChat,Liu2025PersonaFlow,Park2023GenerativeAgents}；JobMate 则只以经历为 grounding，而不预设人格。这使注意力停留在职业轨迹和用户处境上，但参与者有时也认为不同 persona 使用了同一种声音。简单加入虚构人格会削弱来源保真度，也可能使用户把生成的行为归因于原作者。更谨慎的中间路径是，只保留帖子中已经存在的有限风格线索——例如教程式、就事论事或自嘲的语气——明确标注其来自原文，并避免加入没有来源依据的人格特征。这样可以把形式保真度纳入 grounding，同时不把对话模拟等同于帖子作者本人。

\section{Limitations and Future Work}
\label{sec:limitations}

We note six limitations that bound the interpretation of this study and motivate future work.
%% 我们指出六项限制，它们共同界定了本研究结论的解释范围，并为未来工作提供方向。

First, our sample was recruited from a selective university in China and covered three disciplines. These participants may have had more practice articulating discipline-specific experiences than other young job seekers. We observed difficulty with articulation in only a small number of cases, so we cannot determine how common this interactional barrier would be among people with less career knowledge, different educational backgrounds, or greater employment precarity. Future studies should examine more diverse populations and treat the ability to formulate one's situation as a variable rather than a prerequisite.
%% 第一，我们的样本来自中国少数几所选拔性高校，并仅覆盖三个学科。这些参与者可能比其他年轻求职者更熟悉如何表达与专业相关的经历。我们只在少量案例中观察到表达困难，因此无法判断这一互动障碍在职业知识较少、教育背景不同或就业处境更不稳定的人群中有多普遍。未来研究应考察更多样的人群，并把表达自身处境的能力视为需要研究的变量，而不是参与对话的先决条件。

Second, a focused 20--30-minute study session does not reproduce everyday attention or longitudinal career exploration. Participants entered both conditions with an assigned task and researcher presence, whereas ordinary social-media use may be interrupted, incidental, or repeated over time. These differences could change what users notice, disclose, remember, or revisit in either condition. We therefore cannot infer either the size or the direction of the contrast between JobMate and RedNote in everyday use. Longitudinal field studies should examine whether articulated criteria persist, change later searches, or affect subsequent career actions.
%% 第二，一次专注的 20--30 分钟实验无法复现日常注意力或长期职业探索。参与者在研究任务和研究者在场的情况下使用两种系统，而日常社交媒体使用可能是被打断的、偶然发生的，或在较长时间内反复进行的。这些差异可能改变用户在任一条件下注意、披露、记住或重新查看的内容。因此，我们无法推断 JobMate 与小红书在日常使用中差异的大小或方向。长期现场研究应进一步考察，用户明确表达的标准能否持续存在、改变后续搜索或影响之后的职业行动。

Third, JobMate changed several elements at once: structured persona cards, case recommendation, source visibility, retrieval-augmented responses, and SDT-guided conversation. Our comparative design intentionally treated this bundle as a research probe and used RedNote browsing as an analytic contrast. It therefore cannot identify which component caused a particular difference, or whether components interacted. Controlled follow-up studies could vary card structure, retrieval, source presentation, and conversational guidance separately, while preserving the present study's process-oriented questions.
%% 第三，JobMate 同时改变了多个要素：结构化 persona 卡片、案例推荐、来源可见性、检索增强回答和 SDT 引导的对话。我们的比较设计有意把这一组合视为 research probe，并把小红书浏览作为分析性对照。因此，本研究无法确定某一差异由哪个组件造成，也无法判断不同组件如何相互作用。后续控制实验可以分别改变卡片结构、检索、来源呈现和对话引导，同时保留本研究对意义建构过程的关注。

Fourth, evidence about reflection and decision clarity came from self-report, post-session interviews, and observable conversation rather than subsequent career decisions. A more explicit or detailed account does not necessarily indicate a better decision, sustained clarity, or action. Future work could ask participants to produce comparable self-positioning documents before and after use, analyse changes in specificity and considered options, and follow whether these formulations remain meaningful over time.
%% 第四，我们关于反思和决策清晰度的证据来自自我报告、实验后访谈和可观察对话，而不是参与者之后作出的真实职业决策。更明确或更具体的表述并不必然意味着更好的决策、持续的清晰感或实际行动。未来研究可以要求参与者在使用前后分别产出可比较的自我定位文档，分析具体性和所考虑选项的变化，并继续追踪这些表述是否长期保持意义。

Fifth, transforming a public post into a persona changes the mode in which an author's experience is represented. Public availability should not be treated as blanket consent for an AI system to speak in the author's voice or generate claims the author never made. In this study, the interface kept the source post visible and labeled the agent's replies as AI-simulated. These measures do not resolve questions of consent, contextual integrity, attribution, or potential harm. We therefore treat the persona as a model-mediated representation grounded in a post, not as the original author. Any deployment beyond a bounded research setting should establish an appropriate consent and attribution policy, minimise or anonymise identifying information unless its use is authorised, preserve links to source context where ethically appropriate, and clearly separate quoted source material from generated interpretation.
%% 第五，将公开帖子转化为 persona，会改变作者经历被呈现的方式。内容公开可见，不应被视为作者普遍同意 AI 系统以其口吻说话，或生成作者从未表达过的说法。在本研究中，界面持续展示原帖，并把 agent 回应标注为 AI 模拟。但这些措施并不能解决同意、情境完整性、署名或潜在伤害等问题。因此，我们把 persona 视为锚定于帖子的模型中介再现，而不是原作者本人。任何超出限定研究场景的部署，都应建立适当的同意与署名政策；除非获得授权，否则应尽量减少或匿名化身份信息；在伦理上适当时保留原始语境链接；并清楚区分原文引用与生成式解释。

Sixth, the tone-grounding proposal in Section~\ref{subsec:disc-design} has not been implemented or evaluated. Retaining limited stylistic cues from a source post might increase perceptual differentiation across personas, but it could also exaggerate a momentary tone, stereotype the author, or blur the boundary between source and simulation. Future work should compare experience-only and tone-grounded personas while examining source fidelity, attribution, engagement, and users' understanding of what was written by a peer versus generated by the system.
%% 第六，第~\ref{subsec:disc-design}节提出的语气 grounding 尚未被实现或评估。保留原帖中有限的风格线索，可能提高不同 persona 在感知上的辨识度，但也可能夸大某一时刻的语气、对作者形成刻板化呈现，或模糊来源与模拟之间的边界。未来研究应比较仅基于经历与同时保留语气的 persona，并考察来源保真度、归因、参与意愿，以及用户能否理解哪些内容由同伴写下、哪些内容由系统生成。

\section{Conclusion}

We present JobMate, a research probe that transforms authentic peer-authored career posts into structured, source-visible conversational personas. Following a formative study, we conducted a qualitatively led, between-subjects comparative study with 24 participants to examine two ways of engaging with peer career experiences. Native RedNote browsing exposed participants to diverse trajectories but left them to reconstruct cases and interpret comparisons largely on their own. JobMate reorganised this process around case selection and continued dialogue, making previously implicit constraints and self-judgments available for articulation and response. Our central contribution is therefore not evidence that JobMate was generally more effective, but an account of how transforming a peer post into a conversational counterpart can shift career sensemaking from private comparison toward negotiable self-interpretation. When users provided concrete context and the agent offered sufficiently specific interpretations, they could accept, challenge, or revise what those interpretations said about their abilities, experiences, and career fit; opaque matching and generic replies weakened this opportunity. These findings suggest that AI-mediated peer-experience systems should preserve access to original sources, scaffold users' articulation of their circumstances, and keep generated interpretations specific, inspectable, and contestable. In this way, these systems can help users work out what others' experiences mean for them without replacing those experiences or deciding on users' behalf.
%% 我们提出 JobMate：一个将真实同伴撰写的职业帖子转化为结构化、来源可见的对话式 personas 的 research probe。在形成性研究之后，我们开展了一项包含 24 名参与者、以质性分析为主导的组间比较研究，以考察利用同伴职业经历的两种方式。原生小红书浏览让参与者接触到多样的职业轨迹，但主要由他们自行重构案例并解释比较；JobMate 则围绕案例选择和持续对话重新组织这一过程，使此前内隐的个人限制和自我判断能够被表达和回应。因此，我们的核心贡献并不是证明 JobMate 总体上更加有效，而是说明：将同伴帖子转化为对话对象，如何使职业意义建构从私人化的比较转向可协商的自我解释。当用户提供具体情境，并且 agent 给出足够具体的解释时，他们可以接受、质疑或修订这些有关自身能力、经历和职业适配性的说法；匹配依据不透明和回应过于宽泛则会削弱这一机会。这些发现表明，AI 介导的同伴经验系统应保留对原始来源的访问，为用户表达自身处境提供支架，并使生成式解释保持具体、可检查且可质疑。这样，这些系统可以帮助用户理解他人的经历对自己意味着什么，同时不取代这些经历，也不替用户作出决定。

%% ============================================================
%% ACKNOWLEDGMENTS
%% ============================================================
\begin{acks}
[Anonymized for review.]
\end{acks}

%% ============================================================
%% REFERENCES
%% ============================================================
\bibliographystyle{ACM-Reference-Format}
\bibliography{references}
\clearpage
\appendix

\section{Corpus Construction}
\label{app:pipeline}

A four-stage pipeline (Figure~\ref{fig:pipeline}) converts raw posts into conversable personas. We scraped career-related posts from RedNote using biaoda.me, collecting 220 posts for each of three disciplines (Computer Science, Psychology, Chinese Literature); after deduplication, text cleaning, and removal of posts under 100 characters, 414 of 660 remained. We then used GPT-3.5-turbo to filter invalid posts (pure advertisements, low-information and off-topic material) and to classify the rest into \emph{personal experience}, \emph{interview tip}, \emph{industry insight}, and \emph{recruitment information}, categories derived from the information needs formative participants named; manual verification of 100 posts showed 99\% accuracy, leaving 364 valid posts. From personal-experience posts we extracted four structured fields: background info, job seeking outcome, one to three challenges tags, and a one-sentence experience summary; manual verification of 50 extractions showed 94\% accuracy. Finally, posts with a complete narrative arc entered the \emph{Core Persona Pool} (118 personas: CS~39, Psychology~33, Chinese Literature~46), while other stories and the practical categories formed an \emph{Others' Personal Story} corpus and a \emph{Knowledge Base}. Each persona was augmented by dual-track retrieval: Track~1 retrieves the Top-2 peers with similar background and challenges tags, and Track~2 the Top-3 practical resources matching the persona's destination, both by cosine similarity over \texttt{text-embedding-3-small} embeddings.

%% 一条四阶段流水线（图~\ref{fig:pipeline}）把原始帖子转化为可对话的 persona。我们使用 biaoda.me 从小红书爬取求职相关帖子，三个学科（计算机、心理学、中文）各 220 篇；经去重、文本清洗以及剔除少于 100 字的帖子后，660 篇中留下 414 篇。随后我们用 GPT-3.5-turbo 过滤无效帖子（纯广告、低信息量与离题内容），并把其余帖子分为\emph{个人经历}、\emph{面试经验}、\emph{行业见解}与\emph{招聘信息}四类，这些类别源自形成性研究参与者所提出的信息需求；对 100 篇的人工校验显示准确率为 99\%，最终留下 364 篇有效帖子。我们从个人经历帖中抽取四个结构化字段：背景信息、求职结果、一到三个困难标签，以及一句话经历摘要；对 50 条抽取结果的人工校验显示准确率为 94\%。最后，具有完整叙事弧线的帖子进入\emph{核心 persona 池}（118 个 persona：计算机 39、心理学 33、中文 46），其余故事与实用类别则构成\emph{他人经历}语料库与\emph{知识库}。每个 persona 通过双轨检索得到增强：轨道一检索背景与困难标签相似的 Top-2 同伴，轨道二检索与该 persona 去向匹配的 Top-3 实用资源，二者均基于 \texttt{text-embedding-3-small} 向量的余弦相似度。

\section{Implementation}
\label{app:implementation}

The implementation needed to support three research requirements: serve the two conditions from a single corpus, capture interaction data at turn granularity, and keep prompts versioned so that agent behaviour is auditable after the fact. JobMate is a Vue and Node.js web application. Conversational agents use GPT-5.2, while classification and extraction use GPT-3.5-turbo; embeddings use \texttt{text-embedding-3-small} with cosine similarity. We logged every conversation turn together with its persona, retrieved context, and timestamp.
%% 实现需要支持三项研究要求：从单一语料库供给两种条件、以对话轮为粒度采集交互数据，以及对 prompt 做版本管理，使 agent 行为在事后可审计。JobMate 是一个 Vue 与 Node.js 网页应用。对话 agent 使用 GPT-5.2，分类与抽取使用 GPT-3.5-turbo；向量嵌入使用 \texttt{text-embedding-3-small}，并采用余弦相似度。我们记录了每一轮对话及其对应的 persona、检索上下文与时间戳。

\end{document}